\documentclass[12pt]{article}   

\usepackage{graphicx}  
\usepackage{amssymb}
\usepackage{amsmath}
\usepackage{subcaption}
\usepackage{wrapfig}
\usepackage{tabularx, makecell, multirow} 

\usepackage{rotating}
\usepackage{tikz}
\usetikzlibrary{arrows,
                positioning,
                shapes.geometric}

\usepackage{adjustbox}
                
\usepackage{hyperref}
\hypersetup{
    colorlinks=true,
    linkcolor=blue,
    citecolor=blue,
    filecolor= magenta,      
    urlcolor= magenta 
}

\newcommand{\Phit}{\tilde{\Phi}}
\newcommand{\es}{\epsilon_s}
\newcommand{\ep}{\epsilon_p}
\newcommand{\emm}{\epsilon_m}
\newcommand{\ez}{\epsilon_0}
\newcommand{\rr}{{\mathbf r}}

\newcommand{\R}{\mathbb{R}}

\title{VDAC Solvation Free Energy Calculation by a Nonuniform Size Modified Poisson-Boltzmann Ion Channel Model}
\author{Liam Jemison$^1$, Matthew Stahl$^1$, Ranjan K. Dash$^2$, and Dexuan Xie$^1$\thanks{Corresponding author: Dexuan Xie (dxie@uwm.edu).
Address: 1. Department of Mathematical Sciences, University of Wisconsin-Milwaukee, Milwaukee, WI, 53201-0413, USA; 2. Department of Biomedical Engineering, Medical College of Wisconsin, Milwaukee, WI 53226, USA.}}

\normalsize

\date{}

\begin{document}						

\maketitle




\begin{abstract}
The Voltage-Dependent Anion Channel (VDAC) protein is the primary conduit for the regulated passage of ions and metabolites into and out of mitochondria. Calculating its solvation free energy is crucial for understanding its stability, function, and interactions within the cellular environment. In this paper, we introduce a total solvation free energy, $E$, which is the sum of electrostatic, ideal gas, and excess free energies, along with a non-polar energy to yield a zero of $E$ in the absence of charges. We develop numerical schemes for computing $E$ and update the current mesh generation package to accelerate the generation of tetrahedral meshes and improve the quality of meshes for computing $E$. By integrating these schemes and the updated mesh package with our non-uniform size modified Poisson-Boltzmann ion channel (nuSMPBIC), SMPBIC, and PBIC finite element packages, the PDB2PQR package, and the OPM database, we create the VDAC Solvation Free Energy Calculation (VSFEC) package. Using the VSFEC package, we perform comparison tests on the nuSMPBIC, SMPBIC, and PBIC models by using six VDAC proteins and various ionic solutions containing up to four ionic species, including ATP$^{4-}$ and Ca$^{2+}$. We also conduct tests with different neutral voltages and permittivity constants to explore the varying patterns of $E$. Our test results underscore the importance of considering non-uniform ionic size effects and demonstrate the high performance of the VSFEC package in calculating the solvation free energy of VDAC proteins.
\end{abstract}

\section{Introduction}
The Voltage-Dependent Anion Channel (VDAC) protein is the most abundant protein on the outer mitochondrial membrane (OMM) and is the primary conduit for the regulated passage of ions (e.g.~Cl$^-$, Ca$^{2+}$, K$^+$) and metabolites (e.g.~ATP) into and out of mitochondria, thereby playing a critical role in regulating mitochondrial function \cite{colombini2016vdac, colombini2012vdac, lemasters2006voltage, hodge1997regulation, rostovtseva1998atp, rostovtseva1997vdac}. As a major regulatory gateway for ions and metabolites into and out of the mitochondrion, VDAC plays an important role in regulating cell fate (survival/death) \cite{baines2007voltage, camara2017mitochondrial, das2012does, mccommis2012role, pastorino2008regulation, rostovtseva2012vdac, rostovtseva2008tubulin, shoshan2017mitochondrial, shoshan2012vdac, shoshan2010vdac, shoshan2006voltage}. Calculating its solvation free energy is crucial for understanding its stability, function, and interactions within the cellular environment.

Over the past decade, with the availability of high resolution X-ray crystallographic structural data of VDAC \cite{bayrhuber2008structure, hiller2008solution, ujwal2008crystal,zeth2018ten}, significant progress has been made in the modeling and simulation of VDAC using implicit solvent modeling techniques. By treating the water solvent as a dielectric continuum medium, the effect of water molecules on protein simulation is considered implicitly, resulting in one fundamental ion channel modeling tool --- a Poisson-Boltzmann equation (PBE) \cite{honig95,roux99}, which has been useful for a variety of chemical and biological applications \cite{fogolariPoissonBoltzmannEquationBiomolecular2002, chen2011mibpb,luo2002accelerated,xiePBE2013}. However, PBE uses a Boltzmann distribution to model the concentration of ion species, which can produce unrealistic estimates of ionic concentration near strongly charged protein surfaces. Based on the observation that ionic concentration is limited by certain physical constraints, the authors of \cite{borukhovStericEffectsElectrolytes1997} proposed the incorporation of \emph{steric} effects to produce a more realistic modified PBE model in the case of a few simple electrolytes. Further work has extended these size modifications to incorporate more diverse ionic solutions and protein domains \cite{Li_Xie2014b,  nuSMPBE2017,Ying-Xie2015}, and eventually non-uniform ion sizes \cite{dexuanxieEffectiveMinimizationProtocol} and the complex geometry and interfacial conditions arising from ion channel proteins \cite{xieSizeModifiedPoissonBoltzmann2020, xieEfficientFiniteElement2022}. 
Because VDAC modeling involves simultaneous transport and interaction of multiple ionic specie and metabolites, whose sizes are so different that they cannot be simply treated as either volume-less points or balls with the same radius, we recently developed a size modified Poisson-Boltzmann ion channel (SMPBIC) model with each ion having an averaged size \cite{SMPBEic2019} and a non-uniform size modified Poisson-Boltzmann ion channel (nuSMPBIC) model with ions having distinct sizes \cite{nuSMPBEic}. We use the abbreviation PBIC to refer to a Poisson-Boltzmann ion channel model which incorporates no size effects, a special case of the SMPBIC model.  

Like other Poisson-Boltzmann models (e.g. the widely used packages DelPhi \cite{delphi01}, CHARMM \cite{charmm}, and APBS \cite{holst2001}), our SMPBIC and nuSMPBIC models in \cite{SMPBEic2019,nuSMPBEic}  only calculated the electrostatic solvation free energy $E_{es,D_p}$, which is induced by the charges within a protein region, $D_p$, hosting a VDAC protein, ignoring three other basic energy terms as follows:
\begin{enumerate}
    \item The electrostatic solvation free energy $E_{es,D_s}$, which is induced by the charges within a solvent region, $D_s$, in which each ionic concentration function is defined.
    \item The ideal gas solvation free energy $E_{id}$, which accounts for the entropy of an ionic solution when transferring ions from the gas phase to the solvated phase.
    \item The excess solvation free energy $E_{ex}$, which accounts for the free energy required to satisfy ionic size constraint conditions.
\end{enumerate}
 Here ions are assumed to behave like an ideal gas --- they do not interact with each other except through elastic collisions, and the ionic size constraint conditions are given in algebraic equations (see \cite[Eq. (21)]{nuSMPBE2017}). The expressions of these energies are recalled in \eqref{energies-def} in the next section. From their expressions we can see that all the three energies involve triple integrals over $D_s$. Because of the complex molecular structure of a VDAC protein, the geometric shape of $D_s$ becomes extremely irregular and non-convex. Moreover, both $E_{id}$ and $E_{ex}$ involve logarithmic functions of ionic concentrations, and each ionic concentration is required to be positive at every point of $D_s$, causing additional difficulties in their calculation. Overcoming these difficulties requires advanced theoretical and computational techniques. 
 
In this study, we overcome these difficulties by using finite element interpolation techniques and obtain numerical schemes for computing these energies. We then implement these schemes in Python based on the state-of-the-art finite element library from the FEniCS project\cite{fenics-book}, deriving an efficient Python program for computing these energies by a finite element solution of the nuSMPBIC, SMPBIC, or PBIC model.
 
Note that the sum of $E_{es,D_p}$ with $E_{es,D_s}$, $E_{id}$, and $E_{ex}$ gives an electrostatic free energy functional, $F$, whose variation with respect to each ionic concentration leads to the nuSMPBIC model. A nuSMPBIC solution is also known as a minimum point of $F$. Hence, all four free energy terms are important components of the solvation free energy, especially the ideal gas and excess free energies, which account for ionic size effects and the entropy of an ionic solution. Thus, they should be considered in the calculation of VDAC solvation free energy to produce a better description of the solvation process. 

Motivated by the above observations, in this work, we introduce a total solvation free energy, $E$, as a modification of the electrostatic free energy functional $F$ in the expression 
 \[ E=F-E_0 \quad \mbox{ with } \quad F = E_{es, D_p} +  E_{es, D_s} + E_{id} + E_{ex},\]
where $E_0$ denotes the additional energy to reflect the fact that $E$ should be zero in the case without any charges.  
By using a zero value of the electrostatic potential function $u$, we derive an expression of $E_0$ (see \eqref{E0-def} in the next section), whose calculation requires solving a system of nonlinear algebraic equations and computing the volume of the solvent domain $D_s$. In this work, we develop the numerical schemes for computing  $E_0$ and implement them as a Python program.

Clearly, the numerical accuracy of the solvation free energy $E$ strongly depends on the quality of the tetrahedral volume mesh, $D_{s,h}$, of $D_s$. In the nuSMPBIC finite element package, all the related triangular surface meshes and tetrahedral volume meshes including $D_{s,h}$ are generated by ICMPv2 reported in \cite{Chao4mesh2022}, which is the second revision of the ion channel mesh package version 1 (ICMPv1) reported in \cite{liu2015membrane, liu2017quality}. To further improve ICMPv2 in both mesh quality and computer performance, we develop ICMPv3 as one major work of this study. Because of the significant length required to completely describe ICMPv3, we will present this package in another paper. Here we only report one comparison test result, in which ICMPv3 is found to reduce mesh generation CPU time sharply,  from  about 22 seconds (with ICMPv2) to about 5 seconds while improving the mesh quality significantly. Hence, ICMPv3 is selected as the mesh package in the calculation of solvation free energy.

Moreover, we create a  VDAC Solvation Free Energy Calculation (VSFEC) package by integrating ICMPv3 with our Python programs for computing $E$ and the nuSMPBIC, SMPBIC, and PBIC finite element packages. To simplify the usage of the VSFEC package, we write the input programs in Python such that the VSFEC package can download a VDAC Protein Data Bank (PDB) file from the Orientations of Proteins in Membranes (OPM) database at \url{ https://opm.phar.umich.edu} and then transform the PDB file into a PQR file by using the PDB2PQR program package \cite{dolinsky2004pdb2pqr,dolinsky2007pdb2pqr}, automatically, greatly speeding up the calculation of solvation free energy.

Finally, with the VSFEC package, we perform numerical experiments on six VDAC proteins and eight ionic solutions with up to four species. Details on the VSFEC development and the numerical results and discussions are given in the next two sections, respectively, while a summary of this study is presented in the conclusion section. 

\section{Materials and Methods}
\begin{figure}[t]
  \begin{center}
  \vspace{-15mm}
   \includegraphics[width=0.4\textwidth]{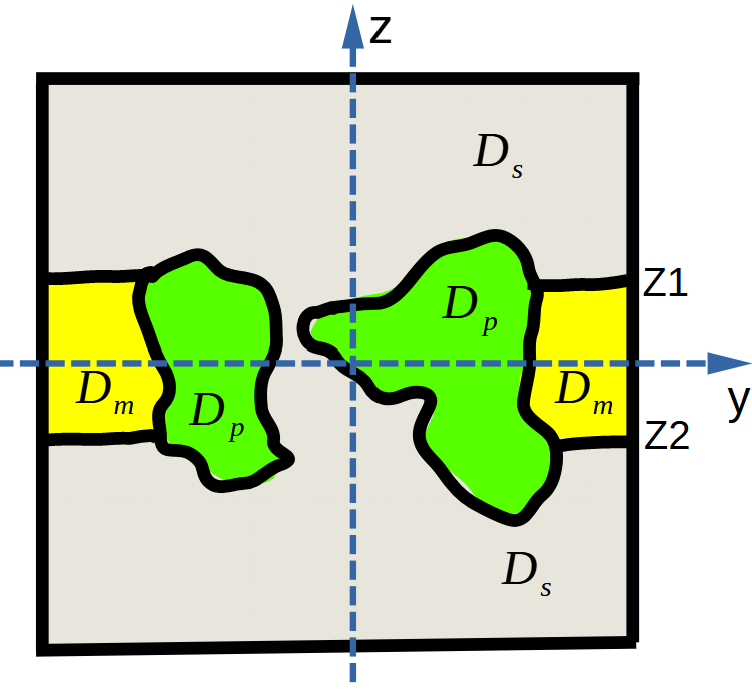}
  \end{center}
  \caption{\small A cross section of the box domain $\Omega$ along $x=0$ as an illustration of the partition \eqref{DomainDecomp}. Here $z=Z1$ and $z=Z2$ define the bottom and top membrane surfaces, respectively, and the origin of the coordinate system is set at the center of the ion channel pore.  }
  \label{boxdomain}
\end{figure}

We define a simulation box domain, $\Omega$,  by
\[ \Omega =\{ (x,y,z) \; | \: L_{x1} < x < L_{x2}, \; L_{y1} < y < L_{y2}, \; L_{z1} < z < L_{z2} \;\},\]
and split it into a protein region, $D_p$, a  membrane region, $D_m$, and a  solvent region, $D_s$, such that 
\begin{equation}\label{DomainDecomp}
\Omega = D_p\cup D_{m} \cup  D_s,
\end{equation}
where $L_{x1}, L_{x2}, L_{y1}, L_{y2}, L_{z1},$ and $L_{z2}$ are real numbers, $D_p$ contains the molecular structure of a VDAC protein with $n_p$ atoms, and $D_s$ contains an ionic solution with $n$ species. An illustration of the box domain partition is given in Figure~\ref{boxdomain}.

Let $u$ denote a dimensionless electrostatic potential function defined in the box domain $\Omega$ and $c_i$ denote the ionic concentration of species $i$ for $i=1,2,\ldots, n$ in  moles per liter (mol/L), defined in the solvent region $D_s$. We derive these functions as a finite element solution of the nuSMPBIC model defined in \cite[Eq. (8)]{nuSMPBEic}. Here $u$ has been split as follows:
\begin{equation}\label{solutionSplit}
u(\rr)=G(\rr) + \Psi(\rr) + \Phit(\rr), \quad \quad   \forall \rr\in \Omega, 
\end{equation}
where $G$ is an analytical expression given in \cite[Eq. (9)]{nuSMPBEic}, $\Psi$ is the solution of a linear interface boundary value problem given in \cite[Eq. (11)]{nuSMPBEic},  and $\Phit$ satisfies a nonlinear system given in \cite[Eq. (12)]{nuSMPBEic}. Here $G$, $\Psi$, and $\Phit$ represent the three electrostatic potentials induced by the atomic charges of an ion channel protein,  membrane charges and Dirichlet boundary and interfaces values, and ionic charges in the solvent region $D_s$, respectively. Since $G$ collects all the singularity points of $u$, $\Psi$ and $\Phit$ can be found numerically without any singularity difficulties. 

\subsection{A total solvation free energy}
A finite element solution of the nuSMPBIC model gives a dimensionless
electrostatic potential function $u$, and $\{c_i\}_{i=1}^n$. Using these $n+1$ functions, we calculate a total solvation free energy, $E$, in  kilocalories per mole (kcal/mol) by the expression 
\begin{equation}
\label{electrostatics-total}
     E  =  E_{es, D_p} +  E_{es, D_s} + E_{id} + E_{ex}  - E_0,
\end{equation}
where $E_{es, D_p}$ and $E_{es, D_s}$ denote the electrostatic solvation free energies within $D_p$ and $D_s$, respectively; $E_{id}$ denotes the ideal gas solvation free energy within $D_s$; $E_{ex}$ denotes the excess solvation free energy within $D_s$; and $E_{0}$ denotes the additional energy to reflect the physical fact that $E$ should be zero if no charge exists in the simulation box $\Omega$. The expressions of $E_{es, D_p}$,$E_{es, D_s}$, $E_{id}$, and $E_{ex}$  have been given in  \cite[Eq. (23)]{nuSMPBE2017}. For completeness, we recall them as follows:
\begin{subequations}
\label{energies-def}
\begin{eqnarray}
& \displaystyle E_{es, D_p} = \frac{ 1}{2} \frac{N_{A}k_B T}{4184} \sum_{j=1}^{n_p} z_j\Phit(\rr_j), &\\ \label{eq:electrostatic-protein}
& \displaystyle E_{es, D_s} = \frac{ 1}{2} \frac{N_{A}k_B T}{4184} \gamma \int_{D_s } u(\rr)   \sum_{i=1}^n Z_i c_{i}(\rr) d\rr, & \\
& \displaystyle E_{id} = \frac{N_{A}k_B T}{4184} \gamma \sum_{i=1}^n \int_{D_s } c_{i}(\rr)\left( \ln \frac{c_{i}(\rr)}{c_i^b} -1\right) d\mathbf{r},  \label{Fes} & \label{Fid}\\
 & \displaystyle E_{ex} = \frac{ 1}{v_0} \frac{N_{A}k_B T}{4184} \int_{D_s } \left[1-\gamma  \sum_{i=1}^n v_i c_i(\rr) \right] 
 \left[ \ln \left(1-\gamma  \sum_{i=1}^n  v_i c_i(\rr)  \right) -1\right] d\mathbf{r}, & \label{Fex}
\end{eqnarray}
\end{subequations}
where $k_B$ is the Boltzmann constant;  $T$ is the absolute temperature; $N_A$ is  Avogadro's number; the constant ${N_{A}}/{4184}$ converts the energy unit from Joules to kcal/mol; $z_j$ and $\rr_j$ denote the atomic charge number and position vector of atom $j$, respectively;  $n$ is the number of ionic species within $D_s$; $Z_i$, $v_i$, and $c_i^b$ are the charge number, ionic volume in cubic angstroms  (\AA$^3$), and bulk concentration (mol/L) of the $i$-th species, respectively; $v_0 = \min_{1\leq i\leq n} v_i$; $\Phit$ is a part of $u$ induced by ionic charges; and $\gamma = N_A /10^{27}$, which is about $6.022 \times 10^{-4}.$ 

Note that the electrostatic free energy within the membrane region $D_m$ has been excluded from the total solvation free energy $E$ of \eqref{electrostatics-total} since it is zero due to the fact that the charge density function within $D_m$ is zero in the nuSMPBIC model as given in the third equation of \cite[Eq. (8)]{nuSMPBEic}. 

\subsection{Expression of the additional free energy}

Obviously, when $\Omega$ does not contain any charge, $G$, $\Psi$, and $\Phit$ become zero, implying that $u=0$, $E_{es, D_p}=0$, and $E_{es, D_s}=0$. As a result, the $n$ algebraic equations of the nuSMPBIC model \cite[Eq. (14)]{nuSMPBEic} can be simplified to a nonlinear system as follows:
\begin{equation}
\label{ci0-eq}
   c_{i}^0 = c_{i}^{b} \left[1-  \gamma \sum\limits_{j=1}^n  v_j c_j^0 \right]^{\frac{ v_i}{v_0}}, \quad i=1,2,\ldots,n,
\end{equation}
where $c_i^0$ denotes a concentration constant of species $i$ corresponding to a zero of $u$. In fact, from \eqref{ci0-eq}, it can be seen that $c_i^0$ is simply the bulk concentration $c_i^b$ when all the ion volumes $v_j=0$. Otherwise, each $c_i^0$ depends on all the ion volumes $v_j$ for $j=1, 2, \ldots, n$. Hence, we need to solve the nonlinear system \eqref{ci0-eq} for these concentration constants. In our notation, $c_i^0$ indicates that $u=0$. For clarity, we refer to $c_i^0$ as the zero-concentration of species $i$.

Since $E=0$ and $c_i = c_i^0$ at $u=0$, from the expressions \eqref{Fid} and \eqref{Fex} of $E_{id}$ and $E_{ex}$, we can get an expression of $E_0$ as follows:
\begin{equation}
\label{E0-def}
    E_0 =  \frac{N_A k_{B}T}{4184} \gamma ||D_s|| \left[ \gamma  \sum_{i=1}^n  c_{i}^0\left( \ln \frac{c_{i}^0}{c_i^b} -1\right)  +
  \frac{1}{v_0} \big(1-\gamma  \sum_{i=1}^n v_i c_i^0 \big) 
 \big[ \ln \big(1-\gamma  \sum_{i=1}^n  v_i c_i^0  \big) -1\big] \right], 
\end{equation}
where $ ||D_s||$ denotes the volume of the solvent region $D_s$. 

In general, we can prove that each zero-concentration, $c_i^0$, defined by the system \eqref{ci0-eq} may differ from the corresponding bulk concentration $c_i^b$.
In fact, if $c_i^0$ is supposed to be $c_i^b$, from \eqref{ci0-eq}, we get 
\[  c_{i}^{b} = c_{i}^{b} \left[1-  \frac{N_A}{10^{27}} \sum\limits_{j=1}^n  v_j c_j^b  \right]^{\frac{ v_i}{v_0}}, \quad i=1,2,\ldots, n,\]
which implies the contradiction,
\[ 1 = \left[1-  \frac{N_A}{10^{27}} \sum\limits_{j=1}^n  v_j c_j^b  \right]^{\frac{ v_i}{v_0}},\]
since $1-  \frac{N_A}{10^{27}} \sum\limits_{j=1}^n  v_j c_j^b$ must be less than one.

\subsection{An iterative scheme for computing zero-concentrations}
 In this section, we present a novel nonlinear iterative scheme for solving the nonlinear system \eqref{ci0-eq} to derive the $n$ zero-concentrations $c_i^0$ for $i=1,2,\ldots, n$. Using this scheme, we only need to solve a series of scalar nonlinear equations instead of a system of $n$ nonlinear algebraic equations. Hence, the calculation of zero-concentrations can be done quickly. 
 
In this scheme, we introduce a new  variable, $p$, defined by 
\[ p=\sum\limits_{j=1}^n  v_j c_j, \]
such that each equation of the nonlinear system \eqref{ci0-eq} is reformulated as
\begin{equation}
\label{ci0-def}
   c_{i} = c_{i}^{b} \big[1-  \gamma p   \big]^{\frac{ v_i}{v_0}}, \quad i=1,2,\ldots,n.
\end{equation}
Multiplying $v_i$ on the both sides of the above equation and summing from 1 to $n$, we can obtain a scalar nonlinear equation in $p$,
\[  p = \sum_{i=1}^n v_ic_i^b \big[1-  \gamma p \big]^{\frac{ v_i}{v_0}}.\]
We then construct a Newton iterative method for solving the above nonlinear equation, numerically, as follows: 
\begin{equation}
\label{Newton4p}
 p^{(k+1)} = p^{(k)} - \frac{g(p^{(k)})}{g^{\prime}(p^{(k)})}, \quad k=0,1,2,\ldots,
\end{equation} 
where $p^{(k)}$ denotes the $k$-th iterate of the Newton method, $g$ and the derivative $g^{\prime}$ of $g$ is given by 
\begin{equation*}
\label{gp-eq}
   g(p) = p - \sum_{i=1}^n v_ic_i^b \big[1-  \gamma p \big]^{\frac{ v_i}{v_0}},
   \quad 
   g^{\prime}(p) = 1 + \frac{\gamma}{ v_0} \sum_{i=1}^n v_i^2c_i^b \big[1-  \gamma p \big]^{\frac{ v_i}{v_0}-1},
\end{equation*} 
 and   $p^{(0)}$ is an initial guess to the solution $p$. By default, we set  
 \[ p^{(0)} = \sum\limits_{j=1}^n  v_j c_j^b. \]
We terminate the iteration and output  $p^{(k+1)}$ as an approximation of $p$ whenever 
\[ |p^{(k+1)} - p^{(k)}| \leq 10^{-10}.\] 
We then get each $c_{i}^0$ by using the formula \eqref{ci0-def}.

\subsection{The SMPBIC and PBIC models}
When each volume size $v_i$ is set as an average size, $\bar{v}$, by   
\[  v_i = \bar{v} \quad \mbox{with} \quad \bar{v}=\frac{1}{n} \sum_{j=1}^n v_j,
     \quad i=1,2,\ldots, n,\]
the nuSMPBIC model is reduced to the SMPBIC model reported in \cite[Eq. (14)]{SMPBEic2019}. In this special case,  we can find $u$ as a solution of a nonlinear interface boundary value problem given in \cite[Eq. (14)]{SMPBEic2019}. We then can get $\{c_i \}_{i=1}^n$ by 
\begin{equation}
 \label{c-selection}
  c_i(\rr) = \frac{c_i^b e^{-Z_{i} u(\rr)} }{1 +  \gamma  \frac{\bar{v}^2}{v_0}   \sum\limits_{j=1}^n c_j^b e^{-Z_{i} u(\rr)} }, \quad \rr \in D_s, \quad i=1,2,\ldots, n.
\end{equation}
Furthermore, we have $v_0=\bar{v}$. Thus, setting $u=0$ in \eqref{c-selection}, we obtain  $c_i^0$ in the explicit expression
\begin{equation}
\label{c0-def-uniform}
   c_{i}^0 =\frac{ c_{i}^{b}}{1 +    \gamma \bar{v} \sum\limits_{j=1}^n  c_j^b}, \quad i=1,2,\ldots,n.
\end{equation}

As another special case, setting all $v_i=0$ can reduce the nuSMPBIC model into the PBIC model reported in  \cite[Eq. (15)]{SMPBEic2019}. In this case, no ionic size effects are considered and the expressions of \eqref{c-selection} are simplified as the classic Boltzmann distribution
\begin{equation}
\label{Boltzmann}
    c_{i}(\rr) = c_{i}^{b}e^{-Z_{i}u(\rr)},    \quad i=1,2,\ldots,n,
\end{equation}
where $u$ is a solution of a nonlinear interface boundary value problem given in \cite[Eq. (15)]{SMPBEic2019}.

In the case of the PBIC model, the excess energy $E_{ex}$ does not exist since none of the ion size constraints are considered. 
Setting $u=0$ in \eqref{Boltzmann} yields
$c_i^0=c_i^b$ for $i=1,2,\ldots,n$. The other energy terms are simplified as follows:
\begin{eqnarray*}
& \displaystyle E_{es, D_p} = \frac{ 1}{2} \frac{N_{A}k_B T}{4184} \sum_{j=1}^{n_p} z_j\Phit(\rr_j), \qquad 
 E_{es, D_s} = \frac{ 1}{2} \frac{N_{A}k_B T}{4184} \gamma \sum_{i=1}^{n} {c}_{i}^b Z_i \int_{D_{s}}   u e^{-Z_{i}u} d\rr, & \\
  & \displaystyle E_{id}= - \frac{N_{A}k_B T}{4184} \gamma \sum_{i=1}^{n} {c}_{i}^b \int_{D_{s}} 
      (Z_i u +1) e^{-Z_{i}u}  d\rr,\qquad 
  E_0 = - \frac{N_{A}k_B T}{4184} \gamma  \|D_s \| \sum_{i=1}^{n} {c}_{i}^b, &
\end{eqnarray*}
Hence, the total electrostatic solvation free energy $E$ for the PBIC model is given in  the expression
\begin{equation}
    \label{electrostatics-total_pbe}
     E = \frac{1}{2} \frac{N_A k_{B}T}{4184} \left( \sum_{j=1}^{n_{p}}z_{j} \Phit(\rr_{j}) - \gamma \sum_{i=1}^{n} {c}_{i}^b \int_{D_{s}} \left[Z_{i}u e^{-Z_{i}u} + 2\left(e^{-Z_{i}u}  -1\right)\right]d\rr \right).
\end{equation}

For a symmetric 1:1 solvent (i.e., $n=2$, $Z_{1}=1$, $Z_{2}=-1$, and ${c}_{1}^b={c}_{2}^b=I_{s}$ with $I_s$ denoting the ionic strength),
the expression \eqref{electrostatics-total_pbe} can be simplified as
\begin{equation*} 
\label{electrostatics-total-salt}
     E = \frac{1}{2} \frac{N_A k_{B}T}{4184} \left(\sum_{j=1}^{n_{p}}z_{j} \Phit(\rr_{j})  + 2 \gamma I_s  \int_{D_{s}} \left[u \sinh(u) - 2\left(\cosh(u)  -1\right)\right] \right)d\rr,
\end{equation*}
which is a modification of the total electrostatic free energy given in \cite{sharp1990calculating,gilson1988calculation}.

{\bf Remark:} The assumption that  $c_i(\rr) \to c_i^b$  as $|\rr|\to \infty$ is  applied to the development of a continuum dielectric solvent model based on the fact  that $u\to 0$  as $|\rr| \to \infty$. This assumption holds in the PBIC model since 
in the PBIC model, each $c_i$ is the Boltzmann distribution \eqref{Boltzmann}; setting $u=0$ in \eqref{Boltzmann}
 immediately gives $c_i(\rr) = c_i^b$ for $\rr\in D_s$. However, when ions are treated as particles with sizes, 
 this assumption should be modified to $c_i(\rr) \to c_i^0$  as $|\rr|\to \infty$.

\subsection{Program package development}
We implemented the expressions of  solvation free energies given in \eqref{electrostatics-total} and our nonlinear iterative scheme for computing the zero-concentrations in Python based on the state-of-the-art finite element library from the FEniCS project \cite{fenics-book}. Here we use a finite element solution of the nuSMPBIC, SMPBIC, or PBIC model and a tetrahedral volume mesh, $D_{s,h}$, of the solvent region $D_s$.

Noting that the calculation of solvation free energies can strongly depend on the mesh quality of $D_{s,h}$, we put significant effort into improving the quality and efficiency of the mesh generation package, ICMPv2  \cite{Chao4mesh2022}, deriving ICMPv3. We then numerically confirmed that ICMPv3 can significantly improve the mesh quality and the computer performance in comparison with ICMPv2. 

To speed up the calculation of solvation free energies, we wrote a Python program to download a VDAC Protein Data Bank (PDB) file from the Orientations of Proteins in Membranes (OPM) database at \url{ https://opm.phar.umich.edu}, instead of from the Protein Data Bank website \url{https://www.rcsb.org}, since each VDAC molecular structure in the OPM database has been transformed to the position required by our nuSMPBIC model. That is, the origin of the coordinate system is set at the center of the ion channel pore with the membrane location numbers $Z1$ and $Z2$ determined by minimizing the transfer energy from water to membrane. 

As required by our mesh program package, we wrote a Python program for transforming the PDB file into a PQR file by using the PDB2PQR program package \cite{dolinsky2004pdb2pqr,dolinsky2007pdb2pqr}. Here a PQR file contains the missed data in the corresponding PDB file including the hydrogen atoms,  the atomic charge numbers, and atomic radii. By default, we generate a PQR file by using the CHARMM force field and removing all the water molecules. We next combined these new Python programs with ICMPv3 and the finite element packages reported in \cite{nuSMPBEic,SMPBEic2019} to derive a VDAC solvation free energy calculation (VSFEC) package. 

With an input Protein Data Bank identifier (PDB ID), we now can use our VSFEC package to download the  PDB file from the OPM database, derive the membrane location numbers $Z1$ and $Z2$, transform the PDB file into a PQR file, generate the  tetrahedral volume meshes and related triangular surface meshes, find a finite element solution of the nuSMPBIC, SMPBIC, or PBIC model, and finally output the solvation free energies. All of these steps can be completed in one run, remarkably speeding up the calculation of solvation free energies. 

\begin{figure}[t]
        \centering
         \begin{subfigure}[b]{0.45\textwidth}
                \centering
               \includegraphics[width=0.7\textwidth]{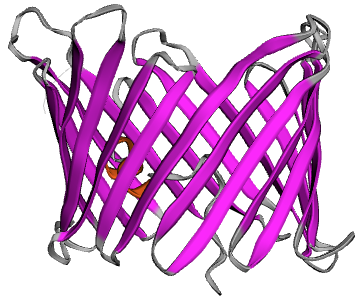}
                \caption{Side view of VDAC}
        \end{subfigure}  
        \qquad
         \begin{subfigure}[b]{0.45\textwidth}
                \centering
               \includegraphics[width=0.7\textwidth]{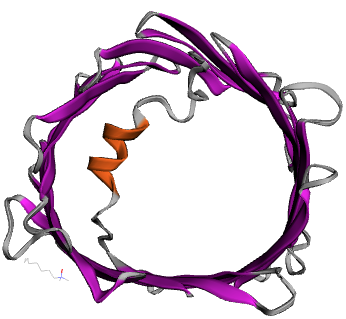}
                \caption{Top view  of VDAC}
        \end{subfigure}
\caption{Two views of a VDAC molecular structure (PDB ID 4BUM) \cite{schredelsekerHighResolutionStructure2014} in cartoon style.} 
\label{fig:4bum-cartoon}
\end{figure}

\section{Numerical Results and Discussion}

In this section, we report the numerical results and discuss them to confirm the importance of considering nonuniform ion size effects in the calculation of solvation free energies, and to demonstrate the high performance of our VSFEC software package. 

We used six different VDAC proteins and eight different ionic solutions with up to four species. The ionic solutions contain two or more of the following species: Cl$^-$,   NO$_3^-$,    Na$^+$,   K$^+$, Ca$^{2+}$, and ATP$^{4-}$. We estimated the ion size of each species by the spherical volume formula with the radius of each sphere given in the website {\em https://bionumbers.hms.harvard.edu/bionumber.aspx?\&id=108517}. From this website, we get the ionic radii of Cl$^-$,   NO$_3^-$,    Na$^+$,   K$^+$, Ca$^{2+}$, and ATP$^{4-}$ as 3.32, 3.35, 3.58, 3.31, 4.12, and 7.0, respectively, in  \AA. Thus, the ion sizes of Cl$^-$,   NO$_3^-$,    Na$^+$,   K$^+$, Ca$^{2+}$, and ATP$^{4-}$ are estimated as 153.29,  157.48,  192.19,  151.91,  292.94, and 1436.76, respectively, in  \AA$^3$.

The six VDAC proteins have  PDB IDs 3EMN\cite{ujwalCrystalStructureMouse2008}, 2JK4 \cite{bayrhuberStructureHumanVoltagedependent2008}, 4BUM \cite{schredelsekerHighResolutionStructure2014}, 5XDN \cite{hosakaCrystalStructuralCharacterization2017}, 6G73 \cite{pdbid6G73} and 6G6U \cite{pdbid6G6U}, respectively.  
Each VDAC is composed of beta-strands with an alpha-helix located horizontally midway within the pore to adopt a beta-barrel architecture. These six VDAC proteins are in the open state and have similar molecular structures. As an example, two views of the molecular structure of the VDAC with the PDB ID 4BUM are displayed in Figure~\ref{fig:4bum-cartoon}.
With these PDB IDs, we got the PDB and PQR files of these six VDAC proteins. We then generated the tetrahedral meshes $\Omega_h$, $D_{p,h}$, $D_{m,h}$, and $D_{s,h}$  for the box domain $\Omega$, protein region $D_p$, membrane region $D_m$, and solvent region $D_{s}$, respectively, with our mesh package ICMPv3. The main data of these proteins and the related box domains and meshes are listed in Table~\ref{mesh_data}. 
All the numerical tests are done on our Mac Studio computer, which has one Apple M1 Max chip and 64 GB memory. For simplicity, we use the default values of the model parameters set in the package. For example, we set $\ep = 2$, $\es = 80$, and $\emm=\ep$ unless otherwise noted. All the related linear finite element equations are solved by the  generalized minimal residual method using incomplete LU preconditioning with the absolute and relative residual error tolerances being  $10^{-8}$. We set the damping parameter $\omega$ introduced in \cite[Eq. (18)]{nuSMPBEic} as  $\omega = 0.1,$ a conservative choice to ensure that our VSFEC package works across a variety of simulations.

\begin{table}[t]
\centering
\scalebox{1}{
  \begin{tabular}{|c|c|c|c|c|c|c|c|c|c|}
   \hline
  \multirowcell{2}{Protein \\PDB ID}& \multicolumn{4}{c|}{Protein data} & \multirowcell{2}{Dimensions of box domain  $\Omega$ \\ $[L_{x_1}$, $L_{x_2}$; $L_{y_1}$, $L_{y_2}$; $L_{z_1}$, $L_{z_2}]$} & \multicolumn{2}{c|}{Number of vertices }\\ \hline
  
&  $n_p$& $n_r$ & $Z1$  & $Z2$   &  &$\Omega_h$&   $D_{s,h}$  \\
\hline
 3EMN&  4,313& 283 & -11.5 & 11.5 & $[-38, 45;  -37, 36; -46, 42]$ &39,305     &  21,632 \\
       \hline
 2JK4 & 4,393 & 288  & -11.7 &  11.7 & $[-41, 44;  -38, 39; -51, 47]$&39,418    & 22,115 \\
    \hline
 4BUM & 4,249& 283 & -11.7 & 11.7 &  $ [-46, 37; -40, 37;  -45, 43]$  &38,879      & 21,614 \\
 \hline
 5XDN & 4,236& 283 & -12.1 & 12.1 &  $ [-39, 38; -37, 38;  -44, 45]$  &38,806      &21,222 \\
  \hline
 6G6U & 4,229& 283 & -11.9 & 11.9 &  $ [-47, 40; -37, 37;  -43, 46]$  &39,251     & 21,642 \\
 \hline
 6G73 & 4,229& 283 & -11.9 & 11.9 &  $ [-49, 38 ; -40, 36;  -43, 48]$  &39,418     & 21,980 \\
        \hline
     \end{tabular}}
  \caption{Protein data, box domain dimensions, and mesh data for the six VDAC proteins used in numerical tests. Here, $n_p$ is the number of atoms,  $Z1$ and $Z2$ are the two membrane location numbers, $n_r$ is the number of residues, $\Omega_h$ denotes a mesh of $\Omega$, and  $D_{s,h}$ denotes a mesh of $D_s$.}
   \label{mesh_data}
\end{table}

\subsubsection{Improvements to mesh generation package}

\begin{figure}[t]
\centering
\text{Tetrahedral meshes produced by our mesh package ICMPv3}
\vspace{3mm}

         \begin{subfigure}[b]{0.32\textwidth}
                \centering
               \includegraphics[width=0.8\textwidth]{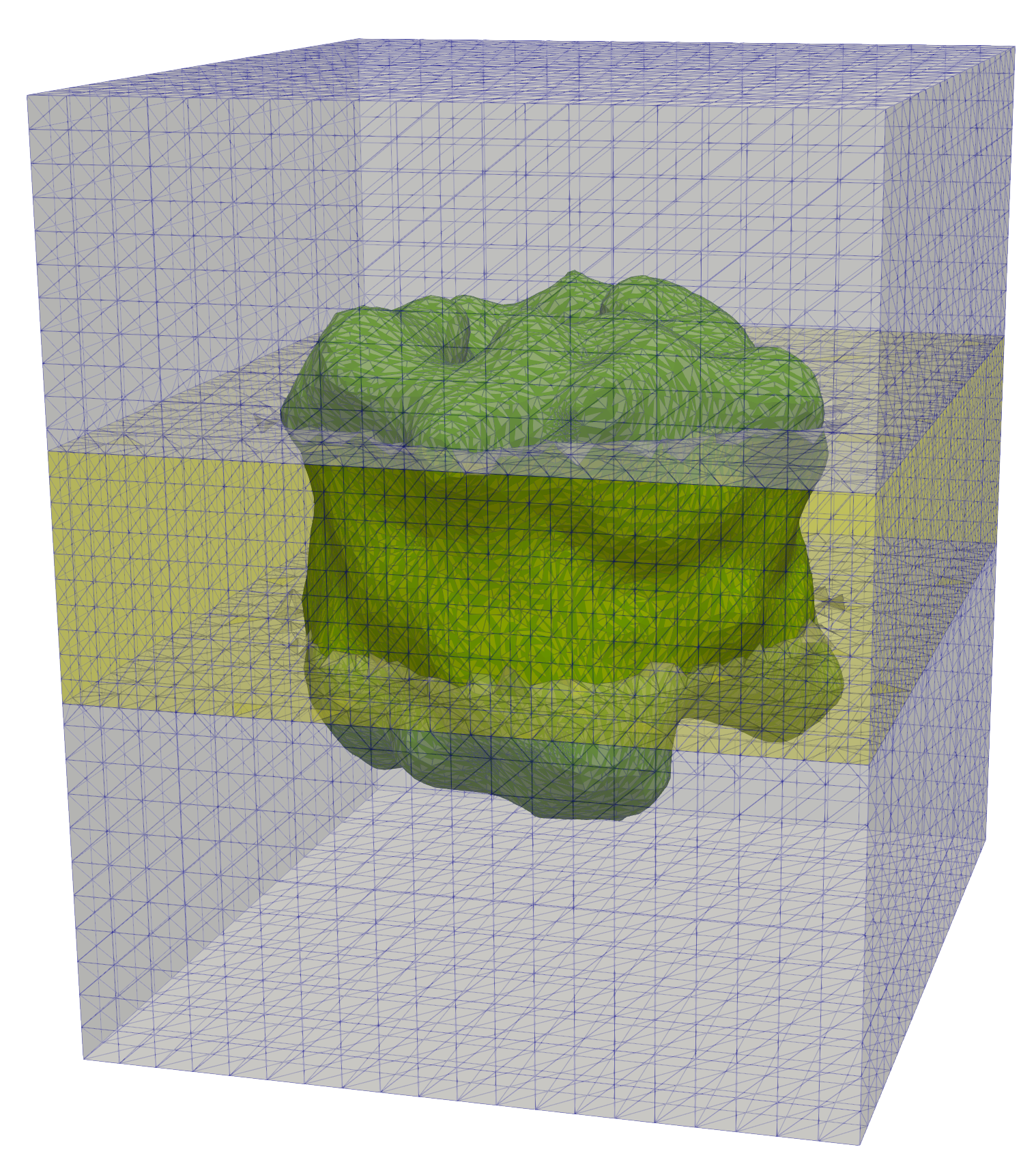}
                \caption{Box domain mesh $\Omega_h$}
               \label{subfig:box-domain-new}
        \end{subfigure}
        \begin{subfigure}[b]{0.32\textwidth}
                \centering
               \includegraphics[width=0.8\textwidth]{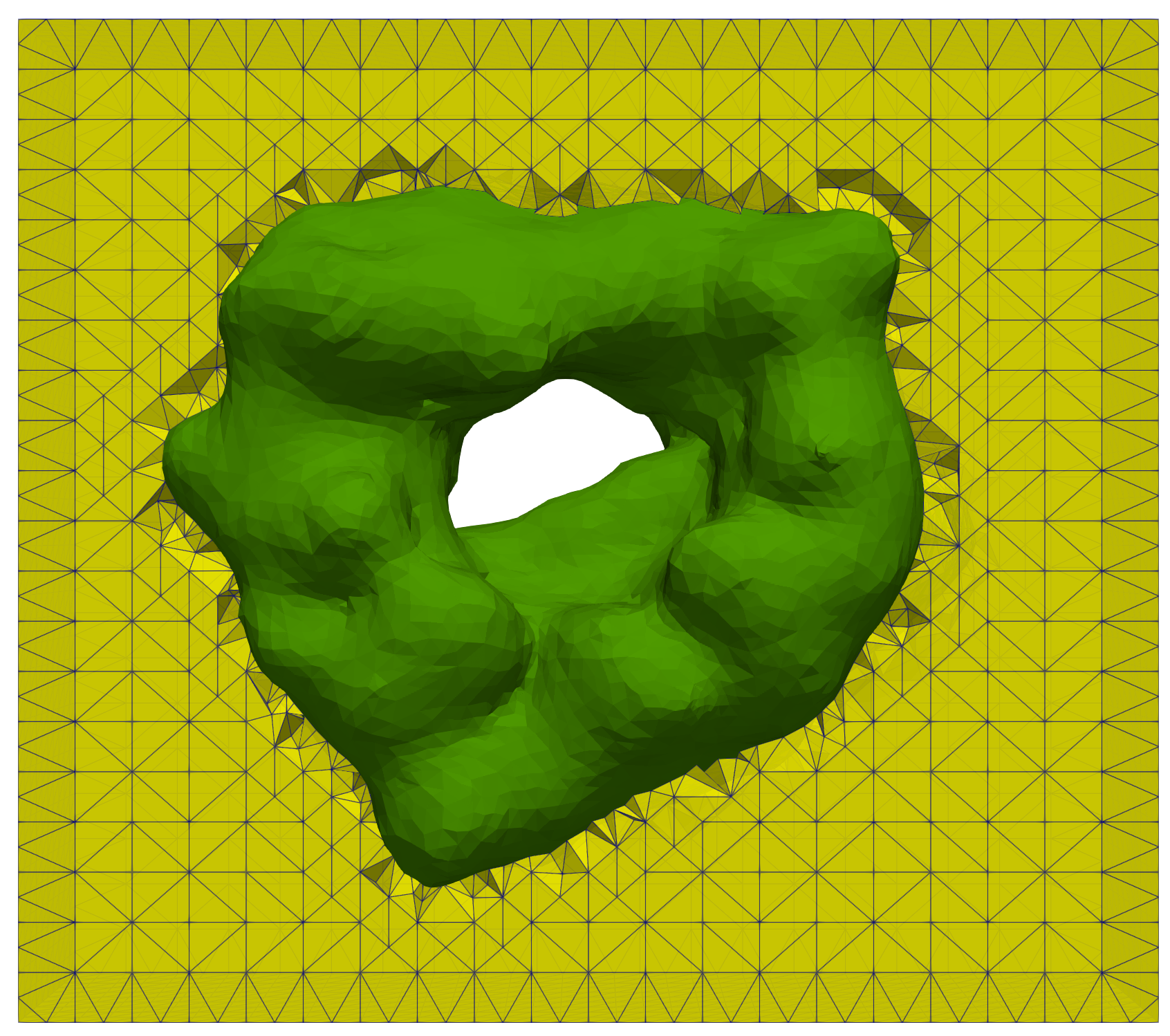}
               \vspace{0.4cm}
                \caption{Top view of $D_{p,h}\cup D_{m,h}$}
               \label{subfig:box-domain-top-new}
        \end{subfigure} 
        \begin{subfigure}[b]{0.32\textwidth}
                \centering
              \includegraphics[width=0.8\textwidth]{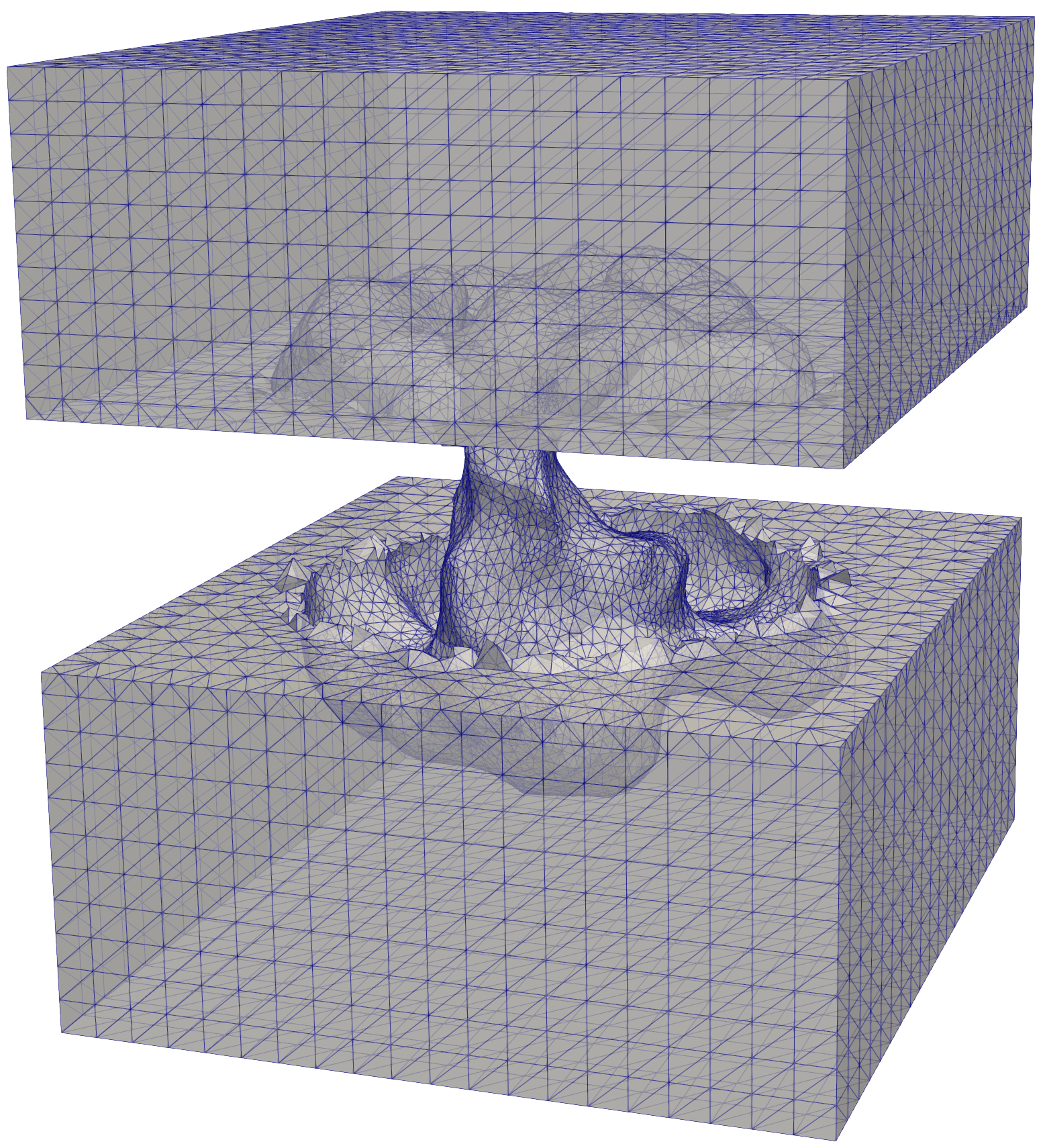}
                \caption{Solvent region mesh $D_{s,h}$ }
               \label{subfig:solvent-domain-new}
        \end{subfigure}\\

       \vspace{3mm}
\text{Tetrahedral meshes produced by the mesh package ICMPv2}
\vspace{3mm}

         \begin{subfigure}[b]{0.32\textwidth}
                \centering
               \includegraphics[width=0.8\textwidth]{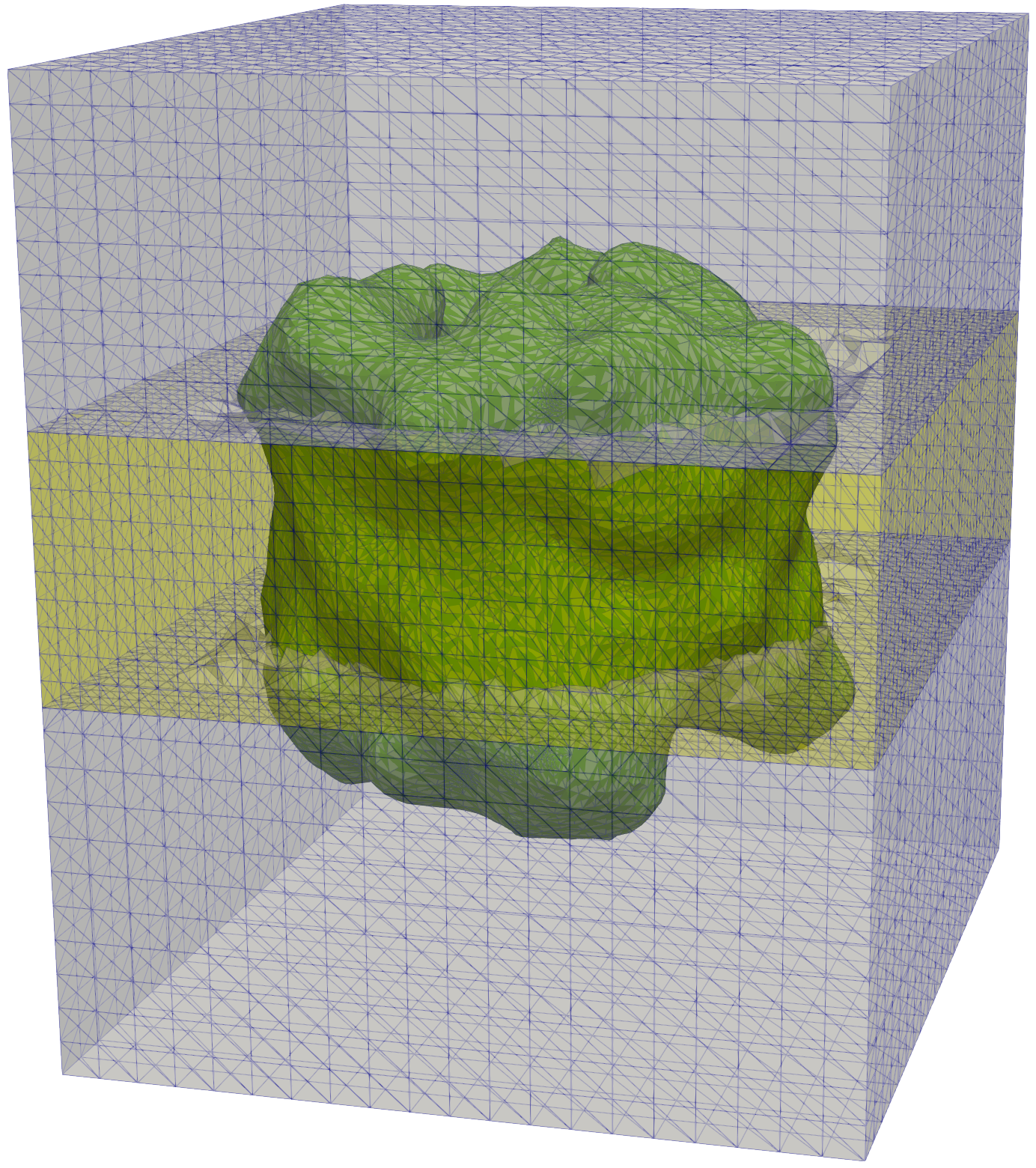}
                \caption{Box domain mesh $\Omega_h$}
               \label{subfig:box-domain-old}
        \end{subfigure}
        \begin{subfigure}[b]{0.32\textwidth}
                \centering
               \includegraphics[width=0.8\textwidth]{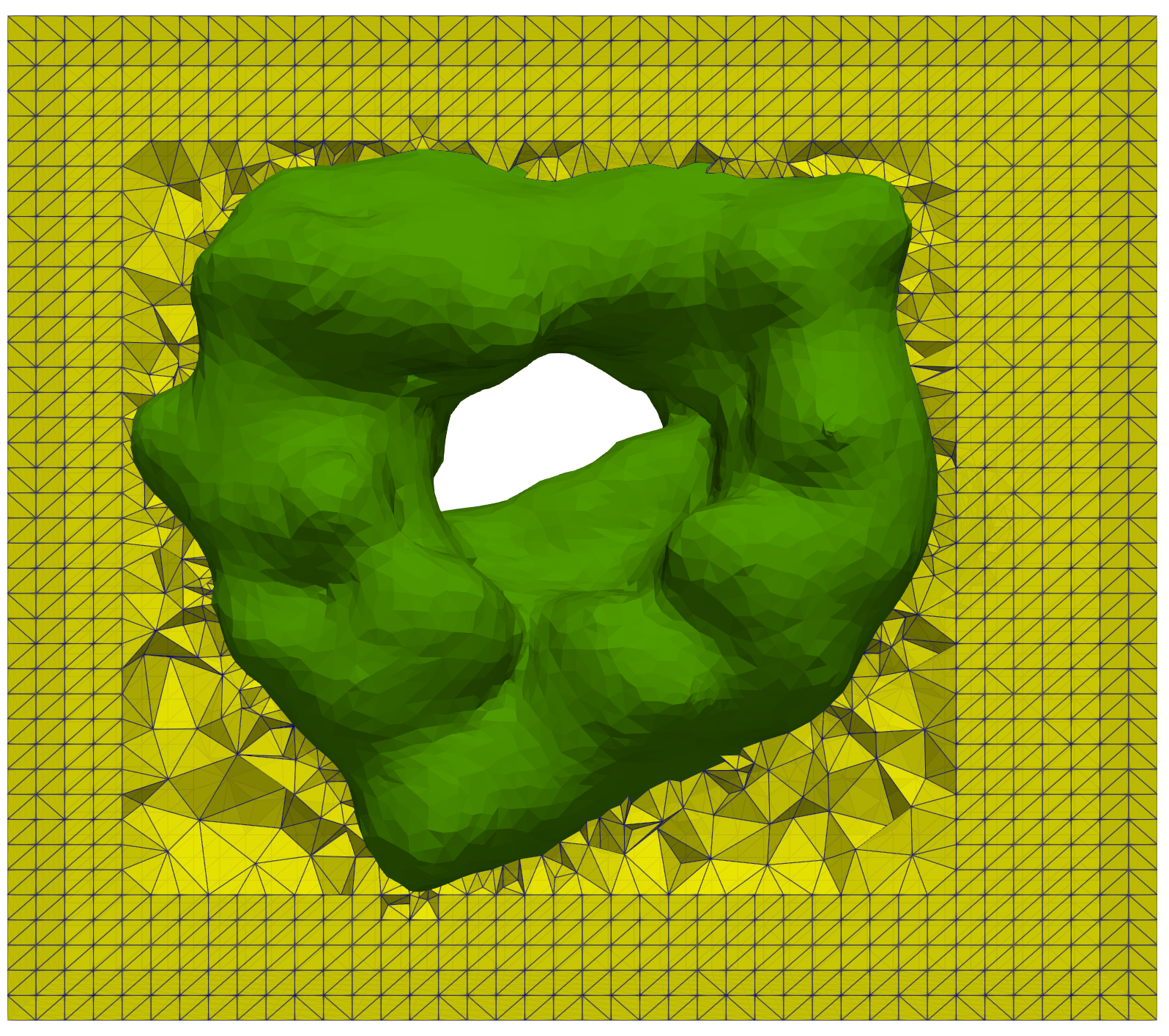}
               \vspace{0.4cm}
                \caption{Top view of $D_{p,h}\cup D_{m,h}$}
               \label{subfig:box-domain-top-old}
        \end{subfigure} 
        \begin{subfigure}[b]{0.32\textwidth}
                \centering
              \includegraphics[width=0.8\textwidth]{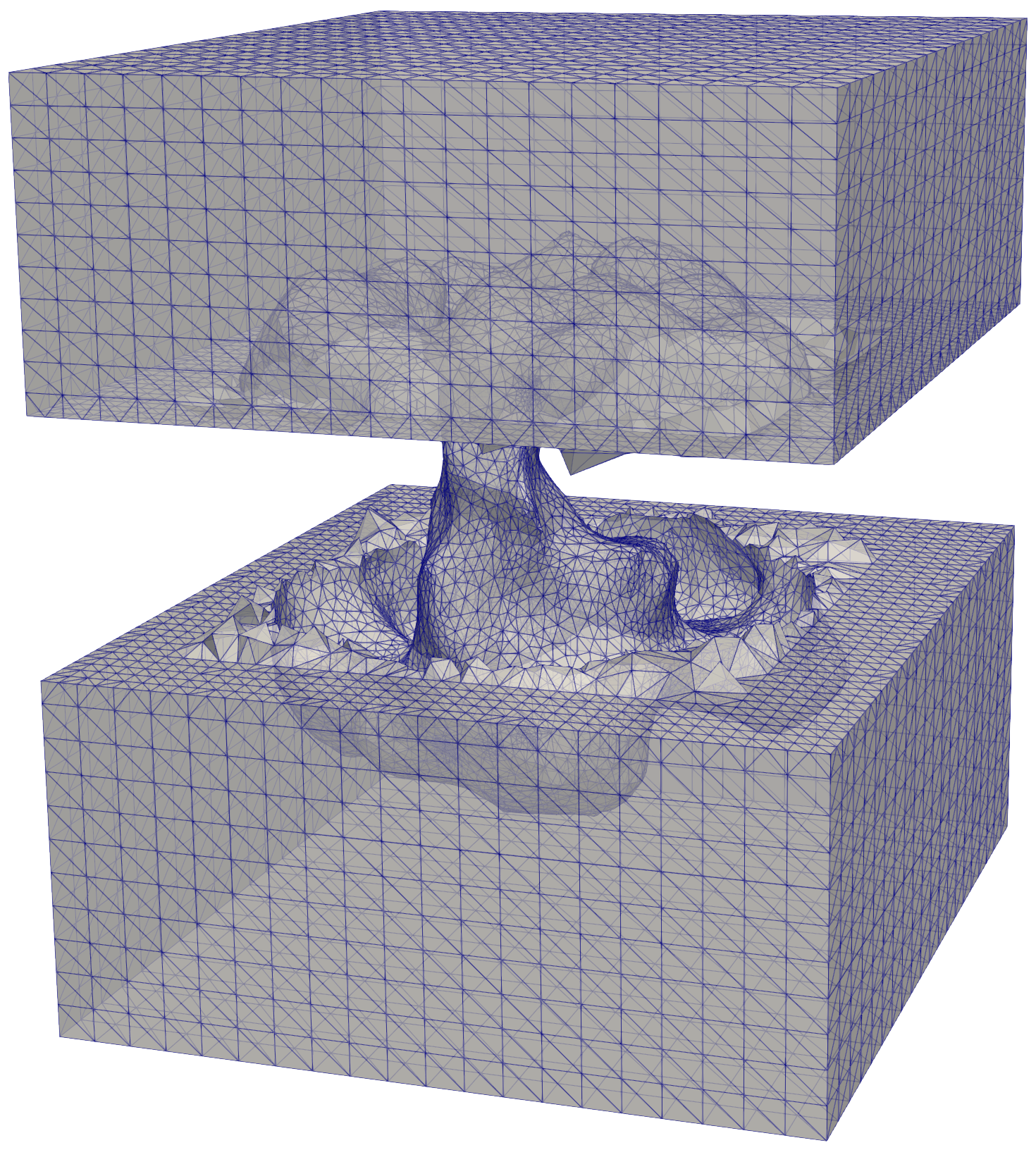}
                \caption{Solvent region mesh $D_{s,h}$ }
               \label{subfig:solvent-domain-old}
        \end{subfigure}
 \caption{A comparison of the interface fitted irregular tetrahedral mesh $\Omega_h$ of the box domain $\Omega$ and the irregular tetrahedral mesh $D_{s,h}$ of  the solvent region $D_s$ generated by our mesh package ICMPv3 with those by the mesh package ICMPv2 \cite{Chao4mesh2022} for the VDAC with PDB ID 3EMN.}       
\label{3EMNstructureMesh}          
\end{figure}

Since all the VDAC proteins have similar molecular structures, their tetrahedral meshes should be fairly similar. Thus, we only report the numerical test results produced by our mesh package ICMPv3 and the current mesh package ICMPv2 for the VDAC protein with PDB ID 3EMN. In Figure~\ref{3EMNstructureMesh}, we compare the meshes produced by ICMPv3 with those produced by ICMPv2, showing that the membrane and solvent meshes generated by ICMPv3 have a higher quality since they fit the protein region much better than those by ICMPv2. These quality improvements are achieved despite a sharp reduction of the mesh generation CPU time. ICMPv2 took about 22.45 seconds while ICMPv3 took only about 5.24 seconds to generate the triangular surface meshes and tetrahedral volume meshes required for computing a finite element solution of the nuSMPBIC, SMPBIC, or PBIC model and the total solvation free energy $E$.
Here the box mesh $\Omega_h$ by ICMPv3 consists of 39,305 vertices and 235,915 tetrahedra while the box mesh $\Omega_h$ by ICMPv2 consists of 39,147 vertices and 232,990 tetrahedra. These test results demonstrate that our mesh package ICMPv3 can significantly improve the mesh quality and computer performance in comparison with the current mesh package ICMPv2. 

Figure~\ref{3EMNstructureMesh} also shows that the box mesh $\Omega_h$ has a complex interface between the solvent region mesh $D_{s,h}$, the protein region mesh $D_{p,h}$ (highlighted in green color), and the  membrane region mesh $D_{m,h}$ (highlighted in yellow color) while the solvent mesh $D_{s,h}$ is extremely irregular and non-convex, which has caused numerical difficulties in the calculation of solvation free energies $E_{id}, E_{es, D_s}$ and $E_{ex}$, given in \eqref{Fes}, \eqref{Fid}, and \eqref{Fex}, which require integrating over $D_{s,h}$ . We overcame the difficulties by finite element interpolation techniques. 

\subsection{Performance of our VSFEC package}
To demonstrate the performance of our VSFEC package in terms of computer CPU time, we did tests on the six VDAC proteins using a mixture of 0.1 mol/L KNO$_3$ and 0.1 mol/L NaCl. The computer CPU times spent on the three major parts of our software package are reported in Table~\ref{tab:runtime}, including the total time, which counts the wall time starting at the input of the parameters until the test job is finished. Thus, the total time includes additional time spent on data input, data output, and mesh generation preprocessing, etc.

Table~\ref{tab:runtime} confirms our package's high performance in terms of computer CPU time in seconds since the total time for each VDAC test took only about between 26 to  80 seconds. In fact, in each test, a finite element solution of the nuSMPBIC model consists of five functions --- one electrostatic potential function, $u$, and four ionic concentration functions, $c_i$ for $i=1, 2, 3, 4$, since the mixture  contains four ionic species. Here $u$ and $c_i$ are defined in the mesh domains $\Omega_h$ with about 39,000 mesh vertices and $D_{s,h}$ with about 22,000 vertices, respectively, resulting in a large system of $N$ nonlinear algebraic equations. Here $N$ is about $39000 + 4\times 22000 = 127000$. From Table~\ref{tab:runtime}, it can be seen that each large nonlinear algebraic system was solved to yield a finite element solution of the nuSMPBIC model in only about 11 to 16 seconds. Our mesh package ICMPv3 also only took about 8 seconds to generate all the tetrahedral volume meshes and triangular surface meshes required for solving the nuSMPBIC model. Moreover, the numerical scheme for computing the solvation free energies is extremely efficient, taking only about 0.2 seconds. Thus, we do not list these time results in Table~\ref{tab:runtime}. These test results confirm that our package is efficient in the calculation of VDAC solvation free energies.

\begin{table}[t]
    \centering
        \begin{tabular}{|c|c|c|c|c|c|c|c|}
    \hline
     Protein&  Generate& \multicolumn{3}{c|}{Calculate $c_1, c_2, c_3, c_4$, and $u$.} & \multicolumn{3}{c|}{Total Time (seconds)} \\\cline{3-8}
    {ID} & Meshes & PBIC & SMPBIC & nuSMPBIC & PBIC & SMPBIC & nuSMPBIC \\\hline
    2JK4 & 8.16 & 33.45 & 10.68 & 12.51 & 49.88 & 27.79 & 29.98 \\ \hline
    3EMN & 7.83 & 62.17 & 10.44 & 13.17 & 78.40 & 26.97 & 30.02 \\\hline
    4BUM & 8.10 & 29.21 & 11.08 & 10.86 & 47.36 & 27.66 & 27.53 \\\hline
    5XDN & 8.27 & 41.38 & 10.75 & 16.45 & 57.71 & 27.60 & 36.07 \\\hline
    6G6U & 7.85 & 41.11 & 10.42 & 12.26 & 57.52 & 27.26 & 27.96 \\\hline
    6G73 & 7.68 & 53.01 & 10.21 & 13.00 & 69.67 & 26.64 & 29.66 \\
    \hline
    \end{tabular}
    \caption{Computer CPU time distributions for computing solvation free energies by our VSFEC package in seconds.
    The total time includes the time spent on mash generation preprocessing and data input-output steps. 
    Here the damping parameter $\omega = 0.4$ except in the case of 5XDN, where $\omega = 0.2$ ensures the convergence of a nonlinear iterative method in the VSFEC package. }
    \label{tab:runtime}
\end{table}

\begin{figure}[t]
 \centering
 \includegraphics[width = 6.4in]{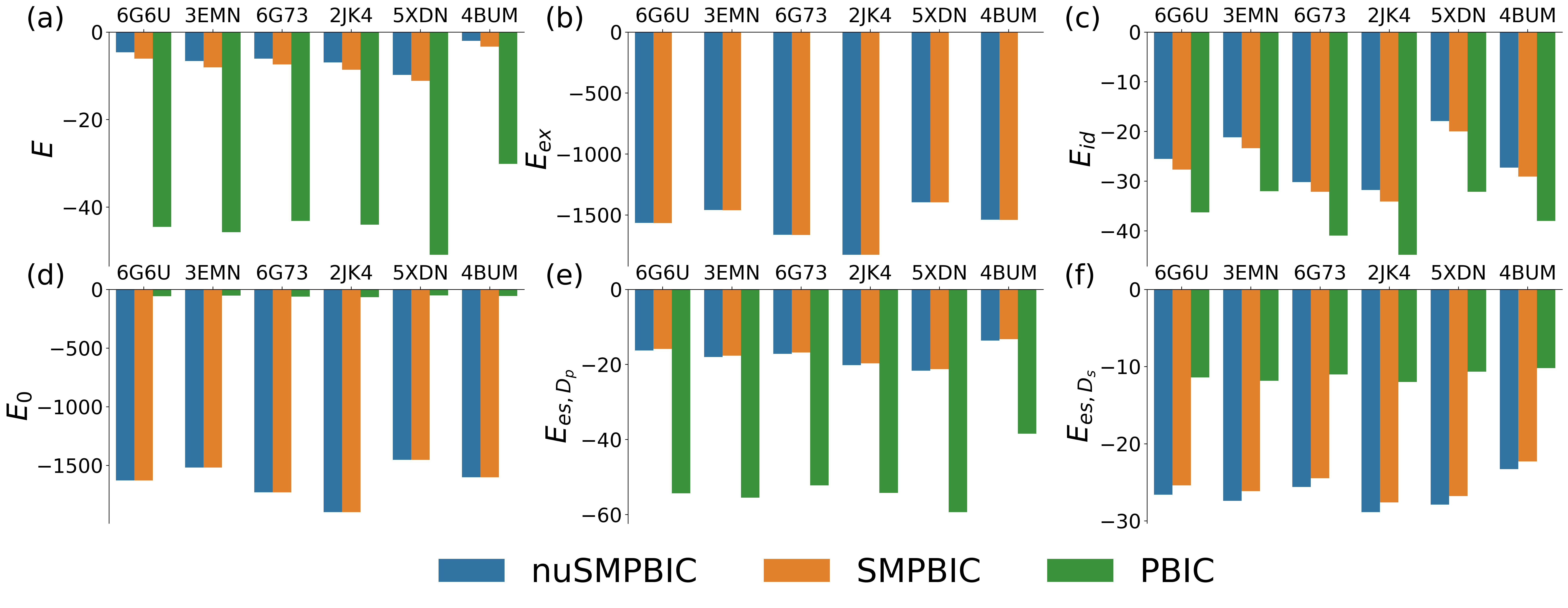}
	\caption{Comparison of the nuSMPBIC model with the PBIC and SMPBIC models in solvation free energy calculation for six VDAC proteins (their PDB IDs marked on the horizontal axes) and a mixture of 0.1 mole KNO$_3$ and 0.1 mole NaCl. }\label{fig:model-comparison} 
\end{figure}

\subsection{Comparison of the nuSMPBIC model with the SMPBIC and PBIC models}

Our software package also includes the SMPBIC and PBIC models since they are the two special cases of the nuSMPBIC model. This enables us to make various numerical experiments for us to compare the nuSMPBIC model with the SMPBIC and PBIC models in the calculation of solvation free energies. We first performed experiments on the six VDAC proteins for a mixture of 0.1 mole KNO$_3$ and 0.1 mole NaCl and reported the calculation results in Figure~\ref{fig:model-comparison}, in which we compare the total solvation free energy $E$, the four component free energies $E_{es, D_p},  E_{es, D_s}, E_{id}$, and $E_{ex}$ defined in \eqref{energies-def}, and the non-polar energy $E_0$ defined in \eqref{E0-def} produced by the three models. 

The magnitude of the total solvation free energy $E$ can vary widely depending on the nature of the solute, the solvent, and the solvation process itself. A negative solvation free energy indicates that the solvation process is thermodynamically favorable, meaning that the solute tends to dissolve in the solvent. A meaningful value of $E$ is typically negative in the range of -5 to -20 kcal/mol. A more negative value of $E$ indicates a stronger tendency for the solute to dissolve in the solvent. For example, a total solvation free energy of $-10$ kcal/mol suggests a more favorable solvation process than that of $-5$ kcal/mol. On the other hand, a positive value of $E$ may indicate that the solvation process is thermodynamically unfavorable, suggesting that the solute may not dissolve easily in the solvent.

Figure~\ref{fig:model-comparison} shows that the nuSMPBIC and SMPBIC models report total solvation free energy values that are significantly smaller in magnitude than the PBIC results while their four component free energies $E_{es, D_p},  E_{es, D_s}, E_{id}$, and $E_{ex}$ are remarkably different, indicating that ionic size effects have a substantial impact on the solvation process. From Figure~\ref{fig:model-comparison} it can also be seen that the nuSMPBIC and SMPBIC models have similar energy values because the sizes of the four ionic species in the solution are close to the average value used in the construction of the SMPBIC model. 

\subsection{Influences of reference state selection}

The electrostatic solvation free energy $E_{es, D_p}$ reflects the energy as required to move a VDAC protein from the reference state to the current state --- the VDAC protein in an ionic solution. It can be calculated as the difference of the current electrostatic solvation energy from the reference electrostatic solvation energy in the expression
\[ E_{es, D_p} =  \frac{N_{A}}{4184} \frac{k_B T}{2e_{c}} \int_{ D_{p}} \rho(\rr) \left[u_{\mbox{\small \em current}}(\rr) - u_{\mbox{\small \em  reference}}(\rr) \right] d\rr, \]
where $u_{\mbox{\small \em current}}$ and $u_{\mbox{\small \em  reference}}$ denote the electrostatic potential functions of the electric fields in the current and reference states, respectively, and $\rho$ is the charge density function in the protein region $D_p$, which is given in the expression:
\[ \rho(\rr) = e_{c}\sum_{j=1}^{n_{p}}z_{j}\delta(\rr-\rr_{j}), \quad  \rr \in D_{p}.\]
Here $\delta_{\rr_{j}}$ is the Dirac delta distribution at $\rr_{j}$.

In this work, we define $u_{\mbox{\small \em  reference}}$ by the Poisson dielectric ion channel model for a VDAC in a  water solvent given in \cite[Eq. 16]{SMPBEic2019} while defining $u_{\mbox{\small \em current}}$ by the nuSMPBIC, SMPBIC, or PBIC model.
With the solution decomposition formula \eqref{solutionSplit}, we get the reference potential $u_{\mbox{\small \em  reference}} = G+\Psi$, resulting in the expression of $E_{es, D_p}$ in \eqref{eq:electrostatic-protein}.

\begin{table}[h!]
    \centering
    \begin{tabular}{|c|c|c|}
    \hline
         protein & $E_{es, D_p}^{vac}$ & $E_{es, D_p}$  \\ \hline 
        \hline
        4BUM & -472.59 & -13.63 \\\hline
        5XDN & -642.68 & -21.68 \\\hline
        3EMN & -545.49 & -18.05 \\\hline
        6G73 & -527.39 & -17.20 \\\hline
        2JK4 & -650.27 & -20.18 \\\hline
        6G6U & -535.84 & -16.27 \\\hline
        \end{tabular}
    \caption{Comparison of electrostatic solvation energy in protein domain computed in reference to a vacuum state and a water state.}
    \label{table:vacuum-comparison}
\end{table}

Traditionally, the quantity  $u_{\mbox{\small \em  reference}}$ is defined by the Poisson equation for a VDAC in the vacuum state as follows:
\[ 
- \epsilon_p \Delta u(\rr) =\alpha \sum\limits_{j=1}^{n_{p}}z_{j}  \delta_{\rr_{j}}, 
  \quad \rr \in \R^3,
\]
where $\alpha = {10^{10}e_{c}^{2}}/{\ez k_{B}T}$ with $\ez$ being the permittivity of the vacuum and $e_c$ the elementary charge, which is about 7042.94. In this case, $u_{\mbox{\small \em  reference}} = G$, which results in the electrostatic solvation energy in the vacuum state, denoted by $E_{es, D_p}^{vac}$, given in the expression
\begin{equation}
    \label{E_ex_vacuum}
    E_{es, D_p}^{vac} = \frac{ 1}{2} \frac{N_{A}k_B T}{4184} \sum_{j=1}^{n_p} z_j \left[\Phit(\rr_j) + \Psi(\rr_j) \right].
\end{equation}

We calculated the electrostatic solvation energies $E_{es, D_p}$ and $E_{es, D_p}^{vac}$ by \eqref{eq:electrostatic-protein} and \eqref{E_ex_vacuum}, respectively, by using a finite element solution of the nuSMPBIC model for the six VDAC proteins and the mixture of 0.1M KNO3 and 0.1M KCl. The results are reported in Table~\ref{table:vacuum-comparison}, from which we can see the values of $E_{es, D_p}$ are significantly lower in magnitude than those of $E_{es, D_p}^{vac}$, falling in a more realistic range.

\begin{figure}[t]
	\begin{subfigure}{0.32\linewidth}
         \includegraphics[width=\linewidth]{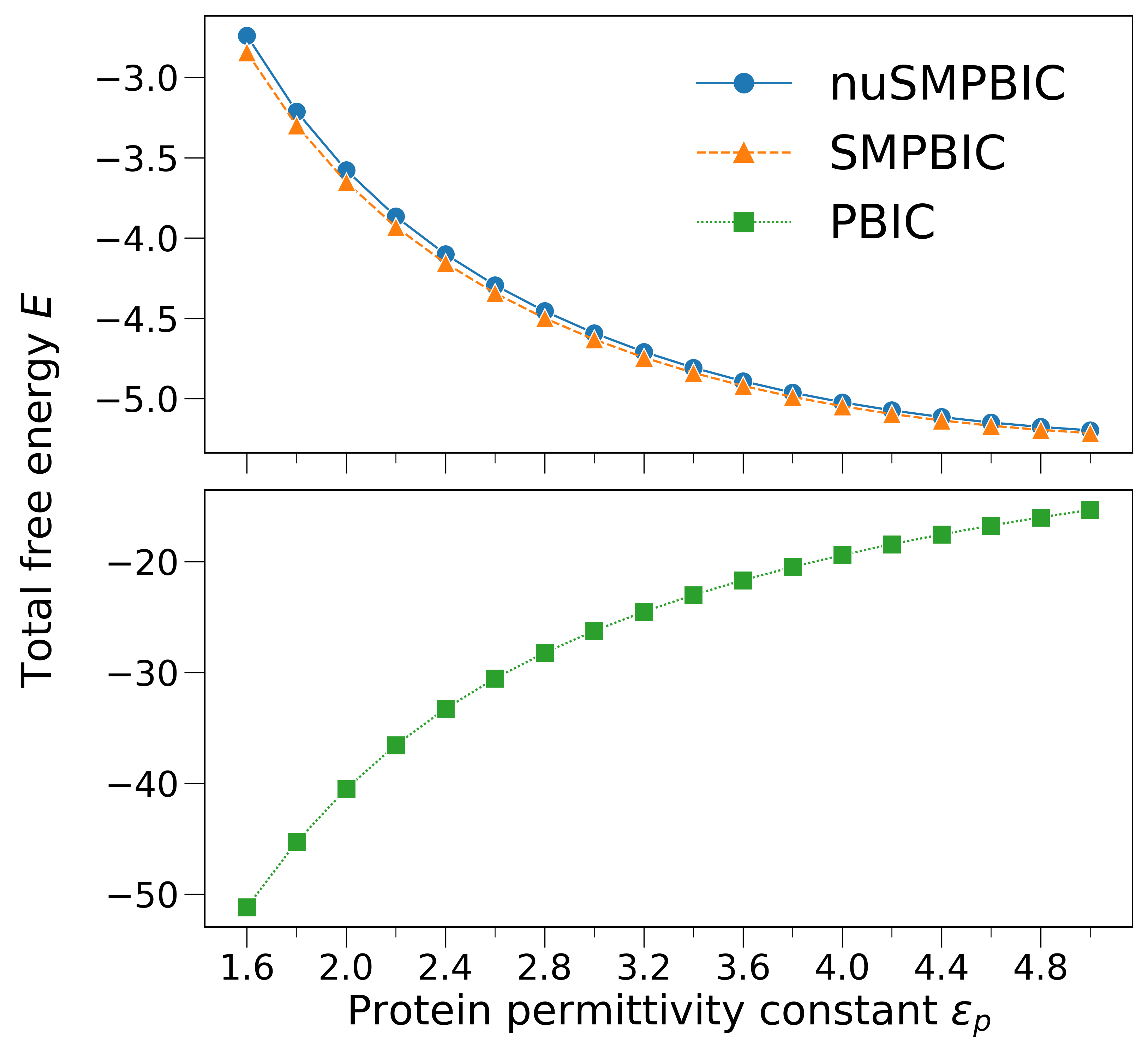}
         \caption{Case of 2 species.}\label{subfig:protein-dielectric-2}
    	\end{subfigure} \hfill
	\begin{subfigure}{0.32\linewidth}
         \includegraphics[width=\linewidth]{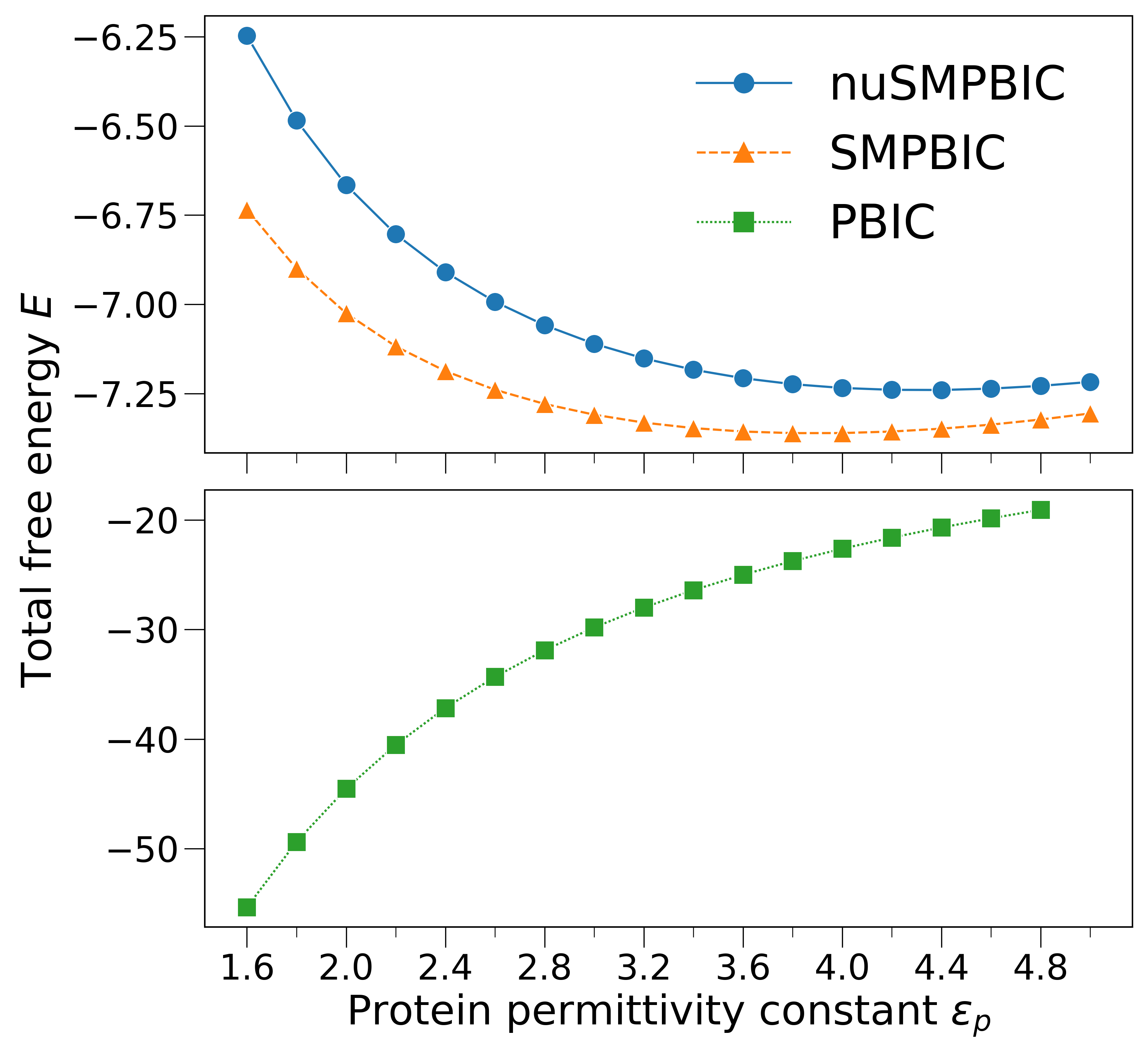}
         \caption{Case of 3 species.}\label{subfig:protein-dielectric-3}
    	\end{subfigure} \hfill
	\begin{subfigure}{0.32\linewidth}
         \includegraphics[width=\linewidth]{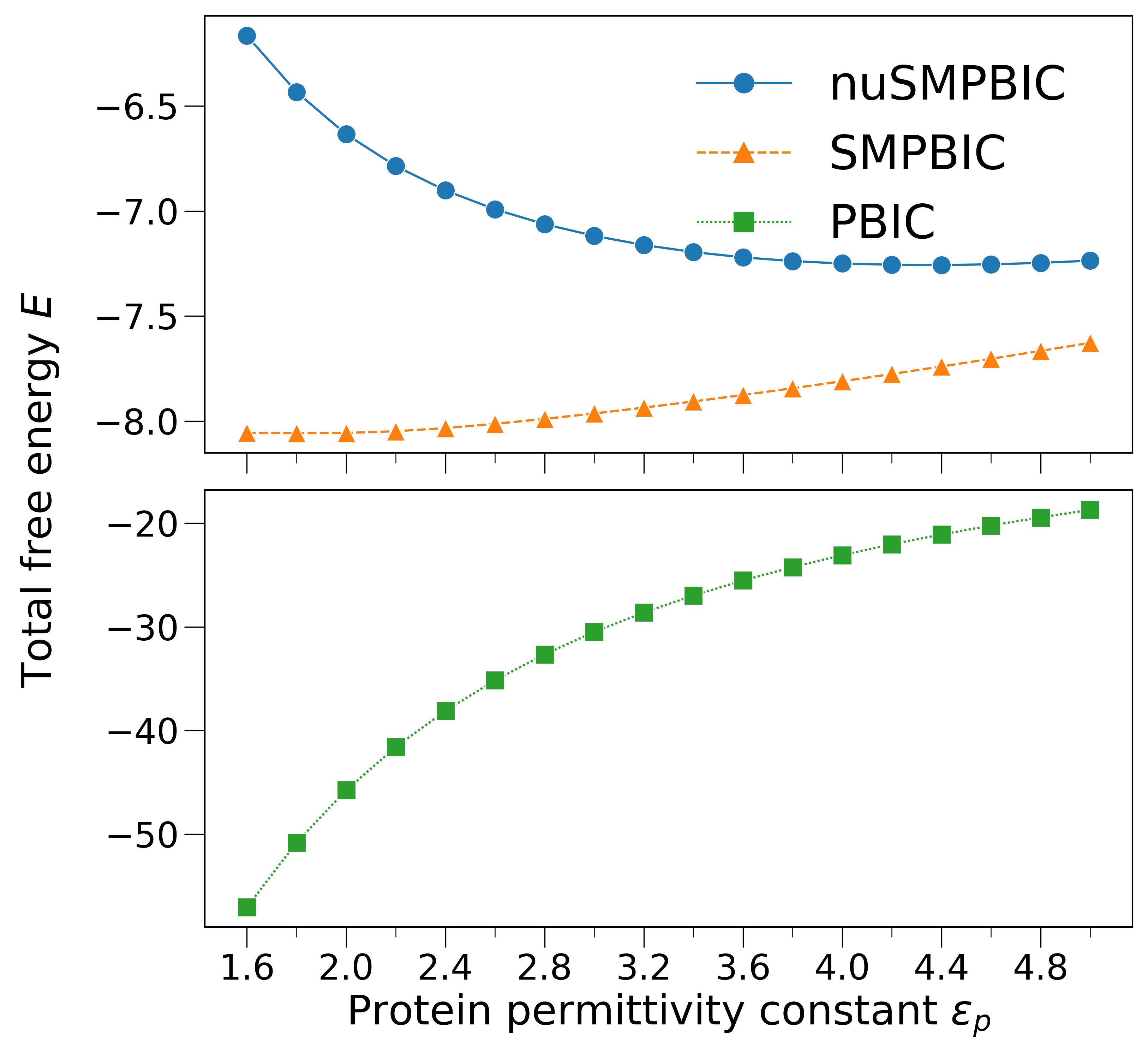}
         \caption{Case of 4 species.}
    	\end{subfigure} \hfill 
	\caption{The total solvation free energy $E$ as a function of protein permittivity  constant $\epsilon_p$. Here we used the default value $\es = 80$ for all the tests. The four species case uses a solution of 0.1 mole $\text{KNO}_3$ and 0.1 mole $\text{NaCl}$,  the three species case uses a solution of 0.1 mole $\text{KNO}_3$ and 0.1 mole $\text{KCl}$, and the two species case uses a solution of 0.1 mole $\text{KCl}$.}\label{fig:protein-dielectric} 
\end{figure}

\begin{figure}
	\begin{subfigure}{0.3\linewidth}
         \includegraphics[width=\linewidth]{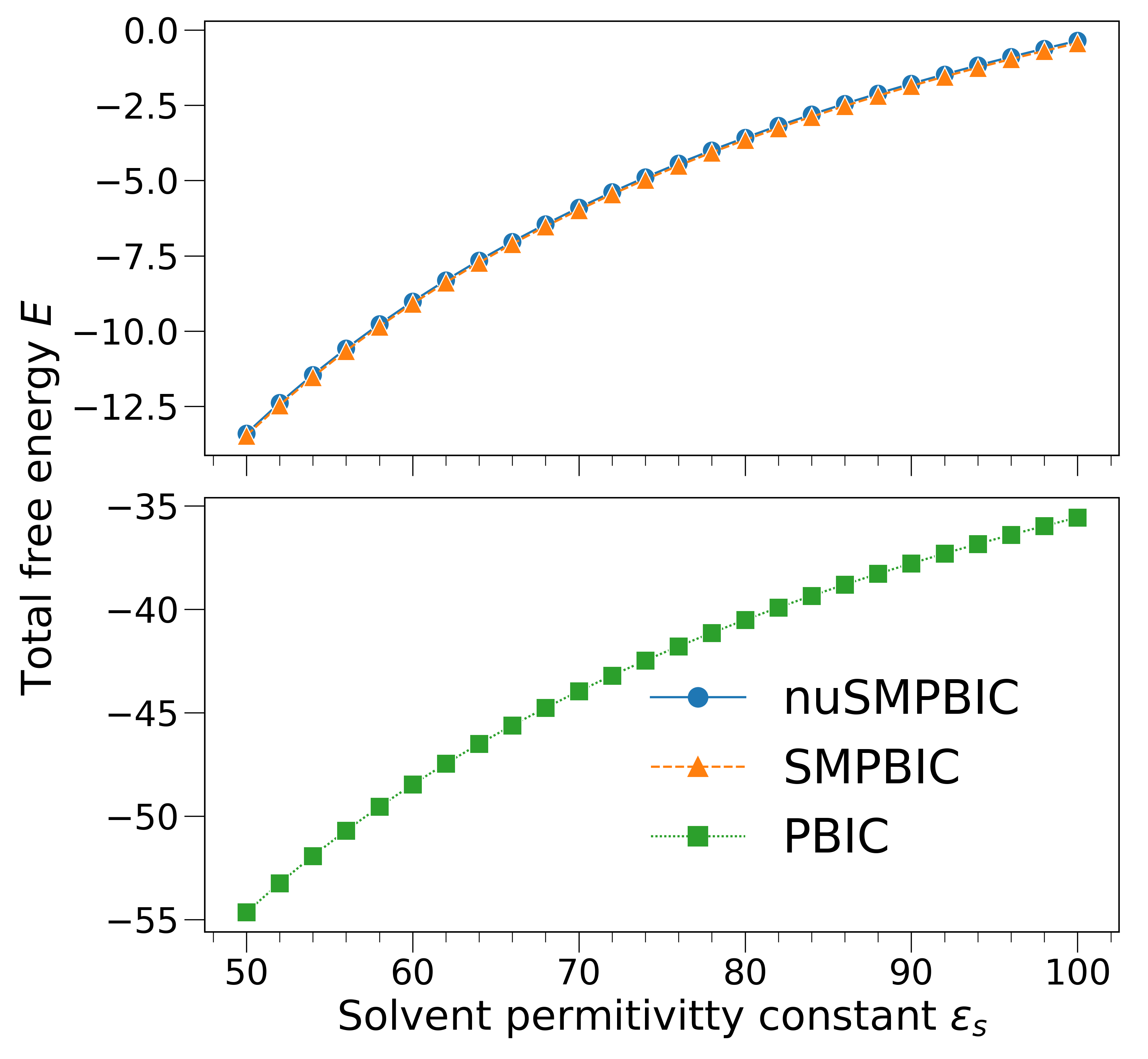}
         \caption{2 ion species case}
    	\end{subfigure} \hfill
	\begin{subfigure}{0.3\linewidth}
         \includegraphics[width=\linewidth]{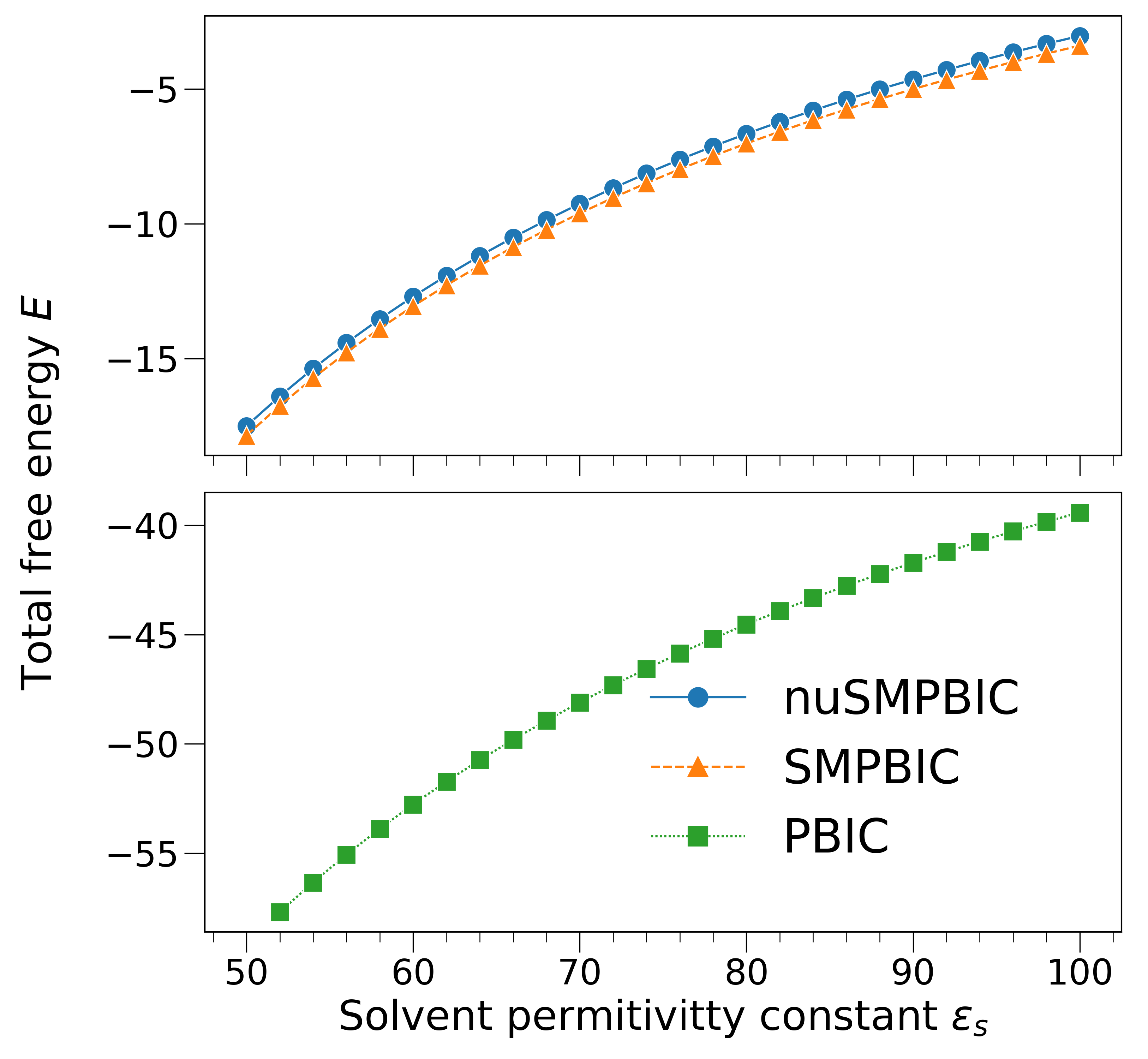}
         \caption{3 ion species case.}
    	\end{subfigure} \hfill
	\begin{subfigure}{0.3\linewidth}
         \includegraphics[width=\linewidth]{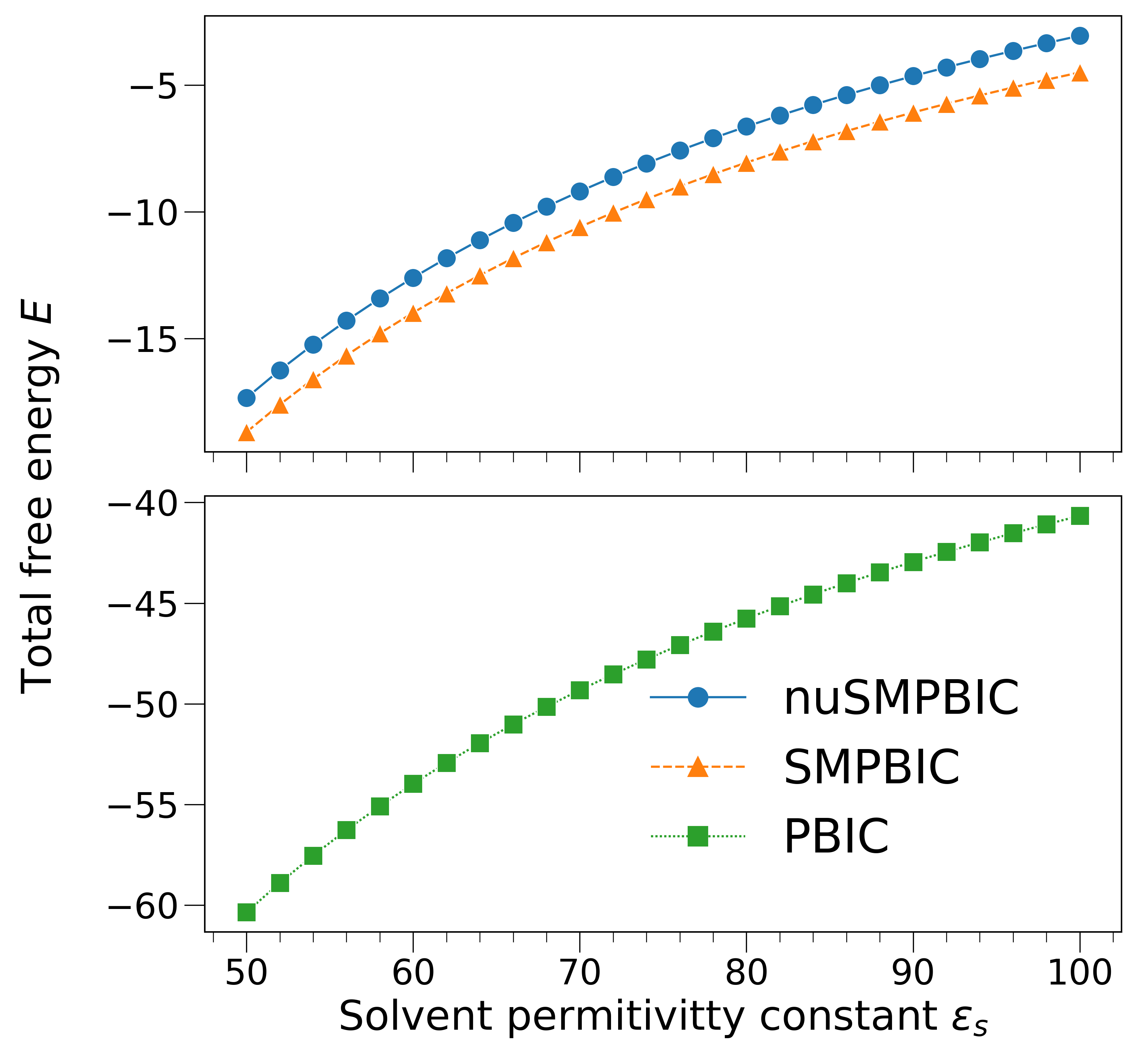}
         \caption{4 ion species case.}
    	\end{subfigure} 
	\caption{The total free energy $E$ as a function of solvent permittivity  constant $\epsilon_s$. Here we used the default value $\ep = 2.0$ for all the tests. The four, three and two ion solutions are the same as those used in figure ~\ref{fig:protein-dielectric}.}\label{fig:solvent-dielectric} 
\end{figure}

\subsection{Influence of protein and solvent permittivity constants}

In these numerical tests, we investigated the influence of the protein and solvent permittivity constants $\ep$ and $\es$ on the calculation of solvation free energies. Test results are reported in Figures~\ref{fig:protein-dielectric} and~\ref{fig:solvent-dielectric}. Here we considered the protein with the PDB ID 3EMN and three different solutions as follows:
\begin{enumerate}
    \item A solution of 0.1 mole $\text{KNO}_3$ and 0.1 mole $\text{NaCl}$  as a case of four species.
    \item A solution of 0.1 mole $\text{KNO}_3$ and 0.1 mole $\text{KCl}$ as a case of three species.
    \item A solution of 0.1 mole $\text{KCl}$ as a case of two species.
\end{enumerate}
 We note that the cases of two and three species do not include sodium, which has a larger hydrated ionic radius (3.58 {\AA}) than potassium, chloride, and nitrate (3.31, 3.32, and 3.35 {\AA}). Correspondingly, from Figures~\ref{fig:protein-dielectric} and~\ref{fig:solvent-dielectric} it can be seen that there is very little difference between the two values of the total solvation free energy $E$ produced by the nuSMPBIC and SMPBIC models, respectively. Figure~\ref{fig:protein-dielectric} shows that varying the protein permittivity constant $\epsilon_p$ in increments of 0.2 between 1.4 and 5 has a significant influence on the value of the total solvation free energy $E$. Plots (a) and (b) in Figure~\ref{fig:protein-dielectric} show that the total solvation free energy $E$ produced by the nuSMPBIC or SMPBIC model is a decreasing function of $\ep$. In contrast, the values of $E$ produced by the PBIC model are increasing with respect to $\ep$. In the four ion species case, where there is more variation in ionic radii, the nuSMPBIC model produced greater values of $E$ than the SMPBIC model. 
 
Figure~\ref{fig:solvent-dielectric} displays the test results of varying the solvent permittivity constant $\epsilon_s$ between 50 and 100 in increments of 2. 
In the four ion species tests, the nuSMPBIC model produces greater free energy values than the SMPBIC model in magnitude. Interestingly, in all the three cases, from Figure~\ref{fig:solvent-dielectric} we can see that $E$ is an increasing function of $\es.$

 Other studies have noted that electrostatic models tend to be sensitive to the choice of dielectric constant and that even when experimental values may be available, properly selecting this parameter is important for obtaining more favorable values of $E$ \cite{aminVariationsProteinsDielectric2020, schutzWhatAreDielectric2001}. Thus, we hope these tests will be illuminating for users of our VSFEC package. The qualitative difference in response of the PBIC model to changes in the protein permittivity constant $\ep$ is of particular note. The monotonic property of the total solvation free energy $E$ with respect to the permittivity constants $\ep$ and $\es$ can be valuable in matching experimental data.

\subsection{Influence of electrostatic potential boundary values}

In our solvation free energy package, a simple Dirichlet boundary value condition is set for the nuSMPBIC, SMPBIC, and PBIC models such that the electrostatic potential function $u$ has the same value, say $u_t$, on the bottom and top surfaces of the box domain $\Omega$. This setting yields a neutral voltage across the membrane as required for a VDAC system to achieve an equilibrium status. The default value of $u_t$ is zero, which was used in the previous tests. 

\begin{figure}[h]
	\begin{subfigure}{0.45\linewidth}
         \includegraphics[width=\linewidth]{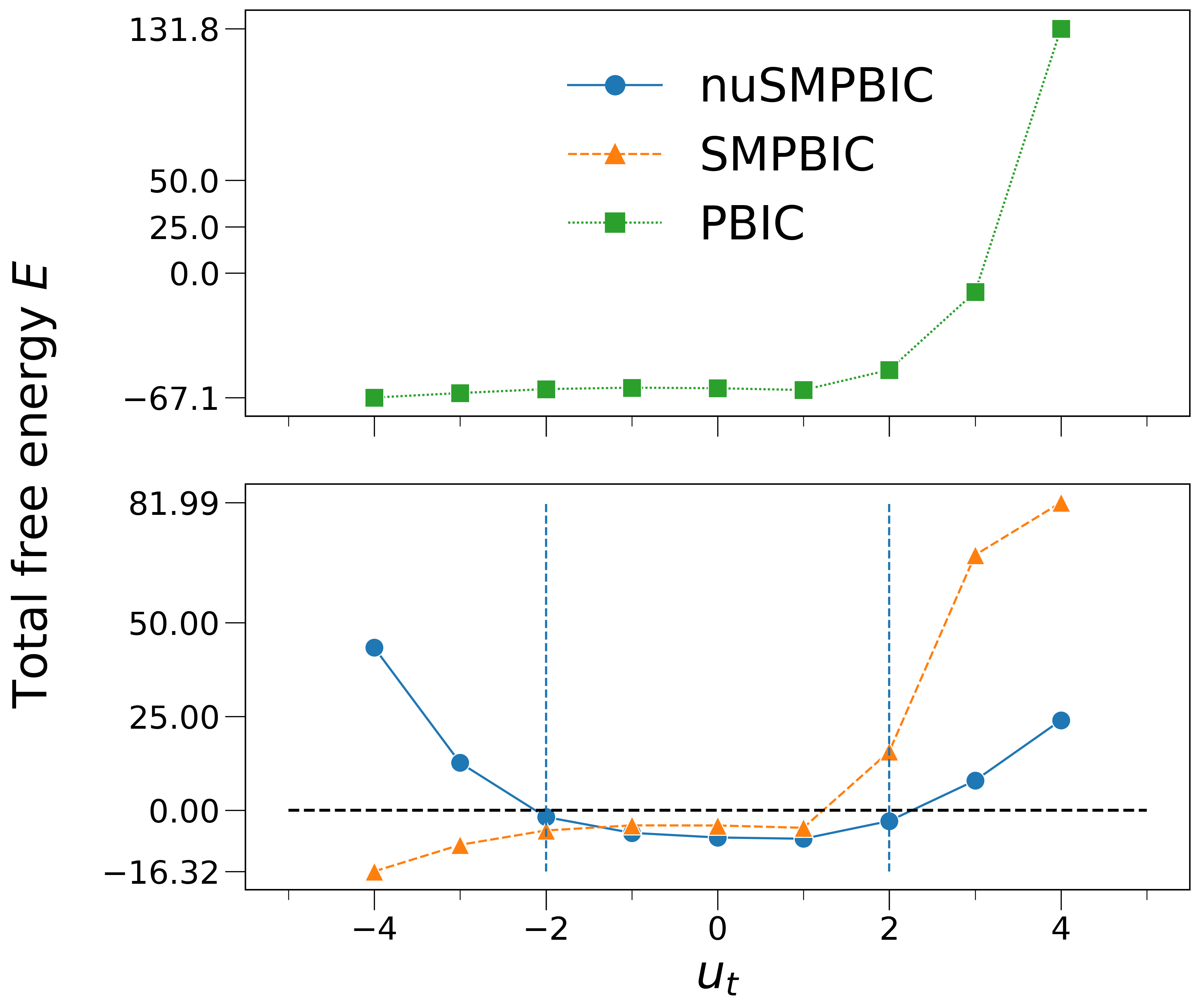}
         \caption{Case of Test 1 with ATP$^{4-}$. }
    	\end{subfigure} \hfill
	\begin{subfigure}{0.45\linewidth}
         \includegraphics[width=\linewidth]{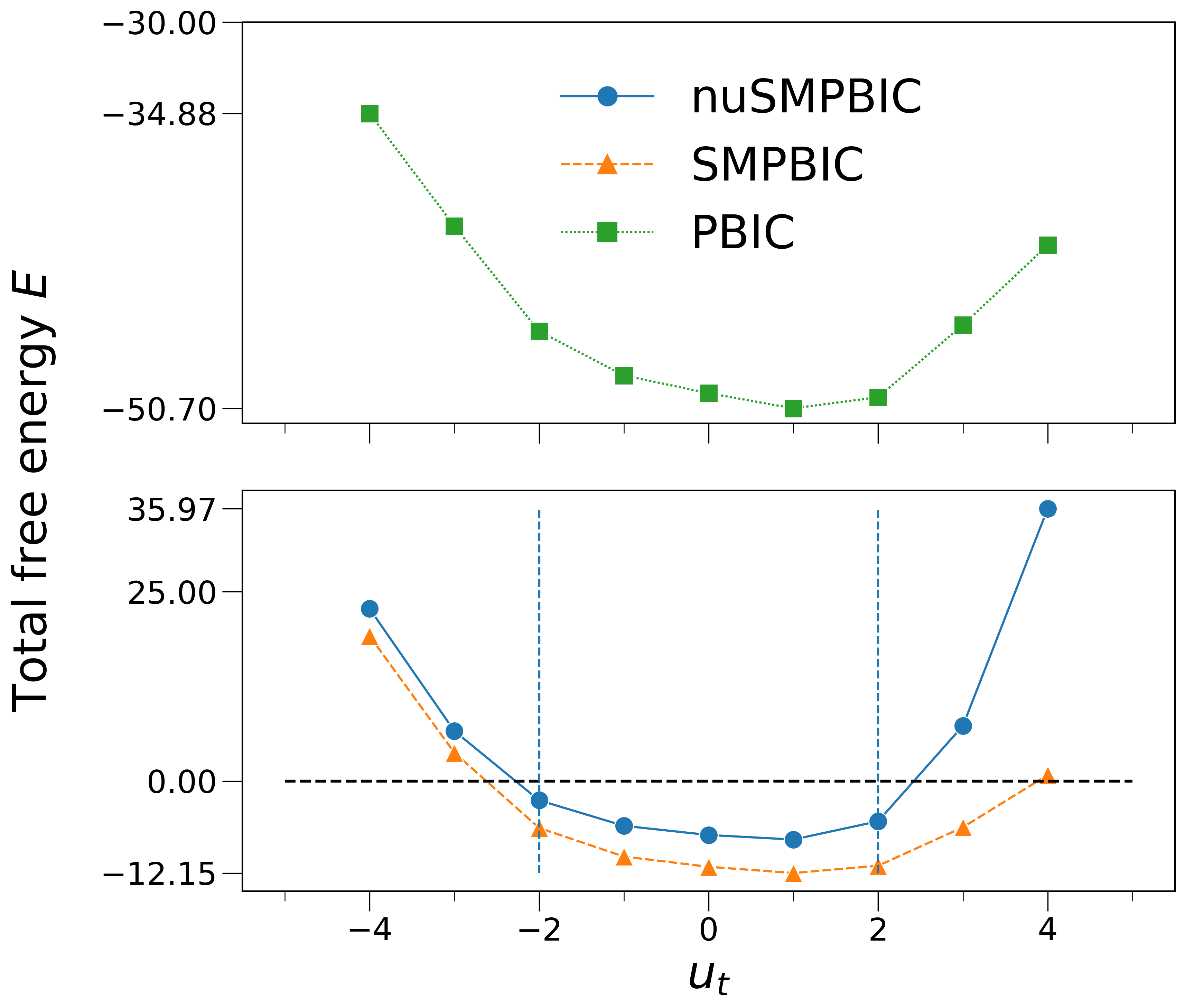}
         \caption{Case of Test 2 with Ca$^{2+}$. }
    	\end{subfigure} 
	\caption{The total solvation free energy $E$ as a function of $u_t$ for $u_t \in [-4, 4]$. Here $u_t$ denotes a value of the electrostatic potential function $u$ on the top side surface of the box domain $\Omega$. In these tests, we set the boundary value of $u$ on the bottom side surface of $\Omega$ to be $u_t$ to maintain a neutral voltage across the membrane as required to achieve an equilibrium status.  }\label{fig:voltage-comparison} 
\end{figure}

With different selections of $u_t$, we calculated the total solvation free energy $E$ by the nuSMPBIC, SMPBIC, and PBIC models for the VDAC protein with the PDB ID 3EMN in two ionic solutions --- a mixture of 0.1M KNO$_3$, 0.1M KCl, and 0.001M ATP (adenosine triphosphate), and a mixture of 0.12M NaCl and 0.04M CaCl$_2$. Here we specify  an atomic radius of 7 angstroms for ATP  using \url{https://bionumbers.hms.harvard.edu/bionumber.aspx?id=106978} as a reference, and each ATP ion has a charge of $-4$. We plotted the values of $E$ as a function of $u_t$ for $u_t\in [-4, 4]$ for each test case in Figure~\ref{fig:voltage-comparison}. 

In the tests using the NaCl and CaCl$_2$ solution, we can see that for all three models, $E$ is a roughly symmetric function of $u_t,$ minimized around $u_t = 0$ or $u_t = 1$ and increasing to the left and right. In all tests, the PBIC model gives a negative values of $E;$ the values of $E$ produced by the nuSMPBIC and SMPBIC models are negative between $[-2, 2].$ 

Interestingly, Figure~\ref{fig:voltage-comparison} indicates that the presence of ATP in a salt solution has remarkably affected the total solvation free energy $E$ of a VDAC protein and increased the complexity of computing solvation free energies. In these tests with the ATP solution, the total free energy curves produced by the SMPBIC and PBIC models show qualitatively different behaviors, while the curve produced by the nuSMPBIC model retains the same symmetry about $u_t = 0$. This reflects the impact of ATP ions on the calculation of solvation free energies due to each ATP ion having a much larger volume size and a much larger charge magnitude than the sodium, chloride and potassium ions. These test results further confirm the importance of considering nonuniform size effects in the calculation of solvation free energies.

\section{Conclusion}

In this paper, we have introduced a total solvation free energy, $E$, as a sum of three basic polar energies (electrostatic, ideal gas, and excess free energies) with one non-polar energy, $E_0$, which makes a zero of $E$ in the absence of charges. A variation of $E$ with respect to each ionic concentration under an electric field can lead to the nuSMPBIC model. Thus, a minimum value of $E$ can be reached at a nuSMPBIC solution. In this sense, our $E$ is one optimal selection for computing the total solvation free energy comprising both polar and non-polar contributions. However, the calculation of $E$ is difficult since it involves triple integrals over an extremely irregular and nonconvex solvent domain $D_s$. Also, the calculation of non-polar energy $E_0$ requires solving a system of nonlinear algebraic equations, producing another difficulty for computing $E$. In this study, we have overcome these difficulties by developing effective numerical schemes. Furthermore, we have updated the current mesh generation package ICMPv2 as ICMPv3, remarkably accelerating the generation of tetrahedral meshes and significantly improving the quality of meshes. Moreover, we have obtained the VDAC Solvation Free Energy Calculation (VSFEC) package as a combination of our new schemes with the updated mesh package, the nuSMPBIC, SMPBIC, and PBIC finite element packages, the PDB2PQR package, and the OPM database.  

Our numerical test results have demonstrated the importance of considering non-uniform ionic sizes in computing solvation free energies and the high performance of our VSFEC package and advanced our understanding of the roles played by the nuSMPBIC, SMPBIC, and PBIC models in the calculation of solvation free energies. They also have confirmed the strong capability of the VSFEC package to handle complex ionic solutions and varying conditions. Thus, our VSFEC package is expected to be a valuable asset for researchers computing VDAC solvation free energies and investigating the complex behavior of VDAC proteins.

\section{Acknowledgement}
This work was partially supported by the National Science Foundation, USA, through the award number DMS-2153376.
 
\bibliographystyle{elsarticle-num}


\begin{thebibliography}{10}
\expandafter\ifx\csname url\endcsname\relax
  \def\url#1{\texttt{#1}}\fi
\expandafter\ifx\csname urlprefix\endcsname\relax\def\urlprefix{URL }\fi
\expandafter\ifx\csname href\endcsname\relax
  \def\href#1#2{#2} \def\path#1{#1}\fi

\bibitem{colombini2016vdac}
M.~Colombini, The {VDAC} channel: molecular basis for selectivity, Biochimica
  et Biophysica Acta (BBA)-Molecular Cell Research 1863~(10) (2016) 2498--2502.

\bibitem{colombini2012vdac}
M.~Colombini, {VDAC} structure, selectivity, and dynamics, Biochimica et
  Biophysica Acta (BBA)-Biomembranes 1818~(6) (2012) 1457--1465.

\bibitem{lemasters2006voltage}
J.~J. Lemasters, E.~Holmuhamedov, Voltage-dependent anion channel ({VDAC}) as
  mitochondrial governator--thinking outside the box, Biochimica et Biophysica
  Acta (BBA)-Molecular Basis of Disease 1762~(2) (2006) 181--190.

\bibitem{hodge1997regulation}
T.~Hodge, M.~Colombini, Regulation of metabolite flux through voltage-gating of
  {VDAC} channels, The Journal of membrane biology 157~(3) (1997) 271--279.

\bibitem{rostovtseva1998atp}
T.~K. Rostovtseva, S.~M. Bezrukov, Atp transport through a single mitochondrial
  channel, {VDAC}, studied by current fluctuation analysis, Biophysical journal
  74~(5) (1998) 2365--2373.

\bibitem{rostovtseva1997vdac}
T.~Rostovtseva, M.~Colombini, {VDAC} channels mediate and gate the flow of
  {ATP}: implications for the regulation of mitochondrial function.,
  Biophysical journal 72~(5) (1997) 1954.

\bibitem{baines2007voltage}
C.~P. Baines, R.~A. Kaiser, T.~Sheiko, W.~J. Craigen, J.~D. Molkentin,
  Voltage-dependent anion channels are dispensable for mitochondrial-dependent
  cell death, Nature cell biology 9~(5) (2007) 550--555.

\bibitem{camara2017mitochondrial}
A.~K. Camara, Y.~Zhou, E.~Tajkhorshid, W.-M. Kwok, Mitochondrial {VDAC1}: A key
  gatekeeper as potential therapeutic target, Frontiers in Physiology 8~(460)
  (2017) 1--18.

\bibitem{das2012does}
S.~Das, C.~Steenbergen, E.~Murphy, Does the voltage dependent anion channel
  modulate cardiac ischemia--reperfusion injury?, Biochimica et Biophysica Acta
  (BBA)-Biomembranes 1818~(6) (2012) 1451--1456.

\bibitem{mccommis2012role}
K.~S. McCommis, C.~P. Baines, The role of {VDAC} in cell death: {F}riend or
  foe?, Biochimica et Biophysica Acta (BBA)-Biomembranes 1818~(6) (2012)
  1444--1450.

\bibitem{pastorino2008regulation}
J.~G. Pastorino, J.~B. Hoek, Regulation of hexokinase binding to {VDAC},
  Journal of bioenergetics and biomembranes 40~(3) (2008) 171--182.

\bibitem{rostovtseva2012vdac}
T.~K. Rostovtseva, S.~M. Bezrukov, {VDAC} inhibition by tubulin and its
  physiological implications, Biochimica et Biophysica Acta (BBA)-Biomembranes
  1818~(6) (2012) 1526--1535.

\bibitem{rostovtseva2008tubulin}
T.~K. Rostovtseva, K.~L. Sheldon, E.~Hassanzadeh, C.~Monge, V.~Saks, S.~M.
  Bezrukov, D.~L. Sackett, Tubulin binding blocks mitochondrial
  voltage-dependent anion channel and regulates respiration, Proceedings of the
  National Academy of Sciences 105~(48) (2008) 18746--18751.

\bibitem{shoshan2017mitochondrial}
V.~Shoshan-Barmatz, S.~De, A.~Meir, The mitochondrial voltage-dependent anion
  channel 1, Ca$^{2+}$ transport, apoptosis, and their regulation, Frontiers in
  oncology 7.

\bibitem{shoshan2012vdac}
V.~Shoshan-Barmatz, D.~Ben-Hail, {VDAC}, a multi-functional mitochondrial
  protein as a pharmacological target, Mitochondrion 12~(1) (2012) 24--34.

\bibitem{shoshan2010vdac}
V.~Shoshan-Barmatz, V.~De~Pinto, M.~Zweckstetter, Z.~Raviv, N.~Keinan,
  N.~Arbel, Vdac, a multi-functional mitochondrial protein regulating cell life
  and death, Molecular aspects of medicine 31~(3) (2010) 227--285.

\bibitem{shoshan2006voltage}
V.~Shoshan-Barmatz, A.~Israelson, D.~Brdiczka, , S.~Sheu, The voltage-dependent
  anion channel ({VDAC}): function in intracellular signalling, cell life and
  cell death, Current pharmaceutical design 12~(18) (2006) 2249--2270.

\bibitem{bayrhuber2008structure}
M.~Bayrhuber, T.~Meins, M.~Habeck, S.~Becker, K.~Giller, S.~Villinger,
  C.~Vonrhein, C.~Griesinger, M.~Zweckstetter, K.~Zeth, Structure of the human
  voltage-dependent anion channel, Proceedings of the National Academy of
  Sciences 105~(40) (2008) 15370--15375.

\bibitem{hiller2008solution}
S.~Hiller, R.~G. Garces, T.~J. Malia, V.~Y. Orekhov, M.~Colombini, G.~Wagner,
  Solution structure of the integral human membrane protein {VDAC}-1 in
  detergent micelles, Science 321~(5893) (2008) 1206--1210.

\bibitem{ujwal2008crystal}
R.~Ujwal, D.~Cascio, J.-P. Colletier, S.~Faham, J.~Zhang, L.~Toro, P.~Ping,
  J.~Abramson, The crystal structure of mouse {VDAC1} at {2.3 $\AA$} resolution
  reveals mechanistic insights into metabolite gating, Proceedings of the
  National Academy of Sciences 105~(46) (2008) 17742--17747.

\bibitem{zeth2018ten}
K.~Zeth, U.~Zachariae, Ten years of high resolution structural research on the
  voltage dependent anion channel ({VDAC}) ---   recent developments and future
  directions, Frontiers in physiology 9 (2018) 108.

\bibitem{honig95}
B.~Honig, A.~Nicholls, Classical electrostatics in biology and chemistry,
  Science 268 (1995) 1144--1149.

\bibitem{roux99}
B.~Roux, T.~Simonson, Implicit solvent models, Biophys. Chem. 78 (1999) 1--20.

\bibitem{fogolariPoissonBoltzmannEquationBiomolecular2002}
F.~Fogolari, A.~Brigo, H.~Molinari, The {{Poisson-Boltzmann}} equation for
  biomolecular electrostatics: A tool for structural biology, Journal of
  molecular recognition: JMR 15~(6) (2002) 377--392.

\bibitem{chen2011mibpb}
D.~Chen, Z.~Chen, C.~Chen, W.~Geng, G.~Wei, {MIBPB}: {A} software package for
  electrostatic analysis, Journal of Computational Chemistry 32~(4) (2011)
  756--770.

\bibitem{luo2002accelerated}
R.~Luo, L.~David, M.~K. Gilson, Accelerated {P}oisson--{B}oltzmann calculations
  for static and dynamic systems, Journal of Computational Chemistry 23~(13)
  (2002) 1244--1253.

\bibitem{xiePBE2013}
D.~Xie, New solution decomposition and minimization schemes for
  {P}oisson-{B}oltzmann equation in calculation of biomolecular electrostatics,
  J. Comput. Phys. 275 (2014) 294--309.

\bibitem{borukhovStericEffectsElectrolytes1997}
I.~Borukhov, D.~Andelman, H.~Orland, Steric {{Effects}} in {{Electrolytes}}:
  {{A Modified Poisson-Boltzmann Equation}}, Physical Review Letters 79~(3)
  (1997) 435--438.


\bibitem{Li_Xie2014b}
J.~Li, D.~Xie, An effective minimization protocol for solving a size-modified
  {P}oisson-{B}oltzmann equation for biomolecule in ionic solvent,
  International Journal of Numerical Analysis and Modeling 12~(2) (2015)
  286--301.

\bibitem{nuSMPBE2017}
D.~Xie, New finite element iterative methods for solving a nonuniform ionic
  size modified {Poisson-Boltzmann} equation, International Journal of
  Numerical Analysis and Modeling 14~(4-5) (2017) 688--711.

\bibitem{Ying-Xie2015}
J.~Ying, D.~Xie, A hybrid solver of size modified {P}oisson-{B}oltzmann
  equation by domain decomposition, finite element, and finite difference,
  Applied Mathematical Modelling 58 (2018) 166--180.

\bibitem{dexuanxieEffectiveMinimizationProtocol}
X.~Dexuan, An {{Effective Minimization Protocol}} for {{Solving}} a
  {{Size-Modified Poisson Boltzman Equation}} for {{Biomolecule}} in an {{Ionic
  Solvent}}, International Journal of Numerical Analysis and Modeling.

\bibitem{xieSizeModifiedPoissonBoltzmann2020}
D.~Xie, S.~H. Audi, R.~K. Dash, A {{Size-Modified
  Poisson}}{\textendash}{{Boltzmann Ion Channel Model}} in a {{Solvent}} of
  {{Multiple Ionic Species}}: {{Application}} to {{Voltage-Dependent Anion
  Channel}}, Journal of computational chemistry 41~(3) (2020) 218--230.

\bibitem{xieEfficientFiniteElement2022}
D.~Xie, An {{Efficient Finite Element Iterative Method}} for {{Solving}} a
  {{Nonuniform Size Modified Poisson-Boltzmann Ion Channel Model}}, Journal of
  Computational Physics 470 (2022) 111556.

\bibitem{SMPBEic2019}
D.~Xie, S.~H. Audi, R.~K. Dash, A size modified {Poisson-Boltzmann} ion channel
  model in a solvent of multiple ionic species: Application to {VDAC}, Journal
  of Computational Chemistry 41~(3) (2020) 218--231.

\bibitem{nuSMPBEic}
D.~Xie, An efficient finite element iterative method for solving a nonuniform
  size modified {P}oisson-{B}oltzmann ion channel model, Journal of
  Computational Physics 470 (2022) 111556.

\bibitem{delphi01}
W.~Rocchia, E.~Alexov, B.~Honig, Extending the applicability of the nonlinear
  {P}oisson-{B}oltzmann equation: Multiple dielectric constants and multivalent
  ions, J. Phys. Chem. B 105 (2001) 6507--6514.

\bibitem{charmm}
B.~R. Brooks, R.~E. Bruccoleri, B.~D. Olafson, D.~J. States, S.~Swaminathan,
  M.~Karplus, {CHARMM}: {A} program for macromolecular energy, minimization,
  and dynamics calculations, J. Comp. Chem. 4 (1983) 187--217.

\bibitem{holst2001}
N.~Baker, D.~Sept, M.~Holst, J.~A. McCammon, The adaptive multilevel finite
  element solution of the {P}oisson-{B}oltzmann equation on massively parallel
  computers, IBM Journal of Research and Development 45 (2001) 427--438.

\bibitem{fenics-book}
A.~Logg, K.-A. Mardal, G.~N. Wells (Eds.), Automated Solution of Differential
  Equations by the Finite Element Method, Vol.~84 of Lecture Notes in
  Computational Science and Engineering, Springer Verlag, 2012.

\bibitem{Chao4mesh2022}
Z.~Chao, S.~Gui, B.~Lu, D.~Xie, Efficient generation of membrane and solvent
  tetrahedral meshes for finite element ion channel calculation, International
  Journal of Numerical Analysis and Modeling 19~(6) (2022) 887--906.

\bibitem{liu2015membrane}
T.~Liu, S.~Bai, B.~Tu, M.~Chen, B.~Lu, Membrane-channel protein system mesh
  construction for finite element simulations, Computational and Mathematical
  Biophysics 1 (2015) 128--139.

\bibitem{liu2017quality}
T.~Liu, M.~Chen, Y.~Song, H.~Li, B.~Lu, Quality improvement of surface
  triangular mesh using a modified laplacian smoothing approach avoiding
  intersection, PLoS One 12~(9) (2017) e0184206.

\bibitem{dolinsky2004pdb2pqr}
T.~Dolinsky, J.~Nielsen, J.~McCammon, N.~Baker, {PDB2PQR: An automated pipeline
  for the setup of {P}oisson--{B}oltzmann electrostatics calculations}, Nucleic
  Acids Research 32~(suppl 2) (2004) W665.

\bibitem{dolinsky2007pdb2pqr}
T.~Dolinsky, P.~Czodrowski, H.~Li, J.~Nielsen, J.~Jensen, G.~Klebe, N.~Baker,
  {PDB2PQR: expanding and upgrading automated preparation of biomolecular
  structures for molecular simulations}, Nucleic Acids Research 35~(suppl 2)
  (2007) W522.

\bibitem{sharp1990calculating}
K.~Sharp, B.~Honig, Calculating total electrostatic energies with the nonlinear
  poisson-boltzmann equation, Journal of Physical Chemistry 94~(19) (1990)
  7684--7692.

\bibitem{gilson1988calculation}
M.~K. Gilson, B.~Honig, Calculation of the total electrostatic energy of a
  macromolecular system: solvation energies, binding energies, and
  conformational analysis, Proteins: Structure, Function, and Bioinformatics
  4~(1) (1988) 7--18.

\bibitem{schredelsekerHighResolutionStructure2014}
J.~Schredelseker, A.~Paz, C.~J. L{\'o}pez, C.~Altenbach, C.~S. Leung, M.~K.
  Drexler, J.-N. Chen, W.~L. Hubbell, J.~Abramson, High {{Resolution
  Structure}} and {{Double Electron-Electron Resonance}} of the {{Zebrafish
  Voltage-dependent Anion Channel}} 2 {{Reveal}} an {{Oligomeric Population}}*,
  Journal of Biological Chemistry 289~(18) (2014) 12566--12577.

\bibitem{ujwalCrystalStructureMouse2008}
R.~Ujwal, D.~Cascio, J.-P. Colletier, S.~Faham, J.~Zhang, L.~Toro, P.~Ping,
  J.~Abramson, The crystal structure of mouse {{VDAC1}} at 2.3 {{{\AA}}}
  resolution reveals mechanistic insights into metabolite gating, Proceedings
  of the National Academy of Sciences 105~(46) (2008) 17742--17747.


\bibitem{bayrhuberStructureHumanVoltagedependent2008}
M.~Bayrhuber, T.~Meins, M.~Habeck, S.~Becker, K.~Giller, S.~Villinger,
  C.~Vonrhein, C.~Griesinger, M.~Zweckstetter, K.~Zeth, Structure of the human
  voltage-dependent anion channel, Proceedings of the National Academy of
  Sciences 105~(40) (2008) 15370--15375.

\bibitem{hosakaCrystalStructuralCharacterization2017}
T.~Hosaka, M.~Okazaki, T.~{Kimura-Someya}, Y.~{Ishizuka-Katsura}, K.~Ito,
  S.~Yokoyama, K.~Dodo, M.~Sodeoka, M.~Shirouzu, Crystal structural
  characterization reveals novel oligomeric interactions of human
  voltage-dependent anion channel 1, Protein Science 26~(9) (2017) 1749--1758.

\bibitem{pdbid6G73}
A.~Razeto, P.~Gribbon, C.~Loew, The dynamic nature of the VDAC1 channels in
  bilayers: human VDAC at 2.7 Angstrom resolution (Apr. 2018).

\bibitem{pdbid6G6U}
A.~Razeto, P.~Gribbon, C.~Loew,  The dynamic nature of the VDAC1 channels in bilayers as revealed by two crystal structures of the human isoform in bicelles at 2.7 and 3.3 Angstrom resolution: implications for VDAC1 voltage-dependent mechanism and for its oligomerization (Apr. 2019).

\bibitem{aminVariationsProteinsDielectric2020}
M.~Amin, J.~K{\"u}pper, Variations in {{Proteins Dielectric Constants}},
  ChemistryOpen 9~(6) (2020) 691--694.
\newblock \href {http://dx.doi.org/10.1002/open.202000108}
  {\path{doi:10.1002/open.202000108}}.

\bibitem{schutzWhatAreDielectric2001}
C.~N. Schutz, A.~Warshel,
 What are the
  dielectric “constants” of proteins and how to validate electrostatic
  models?, Proteins: Structure, Function, and Bioinformatics 44~(4) (2001)
  400--417.


\end{thebibliography}

\end{document}